\documentclass[5p,times]{elsarticle}

\usepackage[numbers]{natbib}
\usepackage{float}
\usepackage{ulem}
\usepackage{graphicx}
\usepackage{url}

\usepackage{array, caption, threeparttable}

\usepackage{subcaption}

\usepackage{xcolor}
\usepackage{stfloats}
\usepackage{tcolorbox}
\usepackage{amsmath}
\usepackage{amsfonts}
\usepackage{multirow}
\usepackage{booktabs}
\usepackage{listings}
\usepackage{pifont}
\usepackage{makecell}
\usepackage{nicematrix}
\usepackage{bm}
\usepackage{threeparttable}

\usepackage{algorithm}
\usepackage{algorithmicx}
\usepackage{algpseudocode}
\usepackage{pgfplots}
\usepackage{hyperref}
\hypersetup{hypertex=true,
colorlinks=true,
linkcolor=blue,
anchorcolor=blue,
citecolor=blue}

\definecolor{verylightgray}{rgb}{.97,.97,.97}

\newcommand{\gao}[1]{\textcolor{black}{#1}}
\newcommand{\tool}{SVA-ICL}

\algnewcommand\algorithmicinitialize{\textbf{Initialize}}
\algnewcommand\Initialize{\item[\algorithmicinitialize]}
\algnewcommand\algorithmicFor{\textbf{For}}
\algnewcommand\for{\item[\algorithmicFor]}

\journal{Information and Software Technology}

\begin{document}

\captionsetup[table]{
  labelfont=bf,
  labelsep=newline,
  singlelinecheck=false,
}

\begin{frontmatter}
	
    \title{{\tool}: Improving  LLM-based Software Vulnerability Assessment via In-Context Learning and Information Fusion}

    \author[NTU]{Chaoyang Gao}
    \ead{gcyol@outlook.com}
       
    \author[NTU,NJU]{Xiang Chen\corref{mycorrespondingauthor}}
    \ead{xchencs@ntu.edu.cn}
     
    \cortext[mycorrespondingauthor]{Corresponding author}
            \author[NTU]{Guangbei Zhang}
    \ead{guangbei0324@gmail.com}
    
    \address[NTU]{School of Artificial Intelligence and Computer Science, Nantong University, Nantong, China}
    \address[NJU]{State Key Lab. for Novel Software Technology, Nanjing University, Nanjing, China}

\begin{abstract}
\textbf{Context:} Software vulnerability assessment (SVA) is critical for identifying, evaluating, and prioritizing security weaknesses in software applications. 

\textbf{Objective:} Despite the increasing application of large language models (LLMs) in various software engineering tasks, their effectiveness in SVA remains underexplored. 

\textbf{Method:} To address this gap, we introduce a novel approach {\tool}, which leverages in-context learning (ICL) to enhance LLM performance. Our approach involves the selection of high-quality demonstrations for ICL through information fusion, incorporating both source code and vulnerability descriptions. For source code, we consider semantic, lexical, and syntactic similarities, while for vulnerability descriptions, we focus on textual similarity. Based on the selected demonstrations, we construct context prompts and consider DeepSeek-V2 as the LLM for {\tool}.

\textbf{Results:} We evaluate the effectiveness of {\tool} using a large-scale dataset comprising 12,071 C/C++ vulnerabilities. Experimental results demonstrate that {\tool} outperforms state-of-the-art SVA baselines in terms of Accuracy, F1-score, and MCC measures. Furthermore, ablation studies highlight the significance of component customization in {\tool}, such as the number of demonstrations, the demonstration ordering strategy, and the optimal fusion ratio of different modalities. 

\textbf{Conclusion:} Our findings suggest that leveraging ICL with information fusion can effectively improve the effectiveness of LLM-based SVA, warranting further research in this direction.

\end{abstract}


\end{frontmatter}

\section{Introduction}
\label{sec:intro}

Software vulnerabilities are weaknesses or flaws in software applications that attackers can exploit, leading to unauthorized access, data breaches, financial losses, operational disruptions, reputation damage, and legal consequences. These vulnerabilities can arise from coding errors, design flaws, and configuration mistakes, and their exploitation can significantly impact organizations and users. To address these risks, software vulnerability assessment (SVA)~\cite{le2022survey,dissanayake2022software} is employed to systematically identify, evaluate, and prioritize security weaknesses within software applications. The primary purpose of SVA is to enhance the security posture of software by proactively detecting and mitigating potential threats, thereby reducing the likelihood of successful attacks, ensuring compliance with regulatory standards, and building trust among users and stakeholders. Recently, researchers primarily used either source code or vulnerability descriptions for SVA. For example, Liu et al.~\cite{liu2019vulnerability} used deep learning methods to classify vulnerability texts and then assessed the severity levels of the vulnerabilities. Le et al.~\cite{le2022use} resorted to vulnerable statements in the code that could be used for software vulnerability assessment.

Large language models (LLMs) have achieved significant success in various software engineering tasks, including code generation~\cite{vaithilingam2022expectation,liu2024your}, vulnerability detection~\cite{purba2023software,zhou2024large}, and source code summarization~\cite{vaithilingam2022expectation,geng2024large}. 
LLMs excel in software engineering tasks due to their ability to harness extensive training, natural language processing capabilities, pattern recognition skills, contextual awareness, and adaptability. Despite their success in these tasks, to the best of our knowledge, the application of LLMs to the SVA task has not been thoroughly investigated in previous studies. 
However, applying LLMs to SVA presents the following challenges. 
First, in-context learning (ICL) has garnered attention for its excellent performance and ease of use without the need for fine-tuning~\cite{dong2022survey}. However, ICL requires the LLM to effectively understand and utilize contextual information from a limited set of demonstrations.
Second, selecting high-quality demonstrations as contextual information for prompt template construction is challenging, as it involves identifying examples that are both relevant and representative of the vulnerabilities being assessed. 
Finally, previous studies usually considered either vulnerability code or vulnerability descriptions~\cite{le2022use,babalau2021severity,hao2023novel}, whereas considering both modalities may allow for a more accurate selection of similar demonstration examples, necessitating an effective information fusion method. Moreover, it is necessary to better evaluate the similarity between the source code and vulnerability descriptions.

To alleviate these challenges, we propose {\tool}, an approach that leverages the in-context learning capabilities of large language models and fuses information from both source code and vulnerability descriptions. We choose ICL because of its ability to be flexibly applied across different tasks without the need for explicit fine-tuning of the large language models. Additionally, ICL can effectively learn and handle tasks using contextual information, even in few-shot scenarios, providing crucial support for improving the adaptability and accuracy of the SVA task. Specifically, {\tool} develops a customized prompt template for SVA. To select high-quality demonstration examples, we employ information fusion, considering the semantic, lexical, and syntactic similarities of the source code, as well as text similarities for vulnerability descriptions. By combining these similarities, we ensure that the selected demonstration examples are highly relevant to the target vulnerabilities, thereby constructing high-quality context prompts.

We evaluate the effectiveness of {\tool} using a dataset containing 12,071 entries related to vulnerabilities. 
In our constructed dataset, we consider the v3 standard of the Common Vulnerability Scoring System (CVSS), which was recently developed for assessing software vulnerability severity more reasonably.
Compared to the CVSS V2 standard, the CVSS V3 standard can provide a more comprehensive, flexible, and accurate framework for assessing and prioritizing vulnerabilities.

We use DeepSeek-V2~\cite{deepseekv2} as the LLM, based on its outstanding performance in handling complex language tasks. It excels not only in processing code information but also shows significant improvements in context window length and processing capability compared to other large language models (such as GPTs). This enables it to better handle long text inputs that include extensive source code and vulnerability descriptions, further enhancing the overall performance of software vulnerability assessment.

Experimental results show that {\tool} outperforms baseline approaches, increasing Accuracy, F1-score, and MCC performance measures by at least 7.68 percentage points (pp), 5.34 pp, and 13.59 pp, respectively. Ablation studies demonstrate the effectiveness of our component settings, such as the number of demonstrations, the demonstration ordering strategy, and the ratio of different information modalities. Our findings indicate that the optimal number of demonstration examples is four, with an ascending order of similarity and a source code to vulnerability description ratio of 70\% to 30\%. These settings help {\tool} achieve the best performance. Additionally, multiple random splits of the dataset confirm our approach's performance improvement consistency when compared with baselines. Finally, we also show the competitiveness of using  DeepSeek-V2 as the LLM when compared to other popular LLMs (such as GPT-3.5 Turbo and GPT-4o).

Our findings suggest that leveraging LLMs for SVA can significantly enhance performance compared to state-of-the-art baselines. 

These promising results highlight the capabilities of LLMs in improving software security and underscore the need for further research in this domain.
Therefore, we call for more studies to explore and refine LLM-based approaches for the SVA task and similar vulnerability-related studies.

\textbf{The novelty and contributions} of our study can be summarized as follows:

\begin{itemize}

    \item \textbf{Perspective.} We are the first to utilize Large Language Models (LLMs) for software vulnerability assessment, introducing a novel approach {\tool}.

    \item \textbf{Approach.} Our proposed {\tool} employs ICL to enhance the LLM’s understanding of the task, utilizing a meticulously designed prompt template. By comprehensively considering both source code and vulnerability descriptions through information fusion, we select high-quality demonstration examples. For source code, we consider semantic, lexical, and syntactic similarities, while for vulnerability descriptions, we focus on text similarity.

    \item  \textbf{Dataset.} We constructed a large-scale dataset (i.e., containing 12,071 entries of source code and vulnerability descriptions) and considered the CVSS v3 standard while previous SVA studies only considered the CVSS V2 standard.

    \item \textbf{Practial Evaluation.} We evaluate {\tool} on our constructed dataset. The results indicate that {\tool} can outperform these baselines. Additionally, ablation experiments confirm the effectiveness of our customization settings in demonstration selection.

\end{itemize}

\textbf{Open Science.} 
To facilitate replication and further research, we share our dataset and source code on GitHub (\url{https://github.com/judeomg/SVA-ICL}).

\textbf{Paper Organization.} 
Section~\ref{sec:Background} introduces the research background and challenges of applying LLMs to SVA.
Section~\ref{sec:Framework} shows the framework of our proposed approach {\tool} and its details. 
Section~\ref{Experimental Setup} describes our experimental setup, including research questions, experimental subjects, baselines, performance measures, and implementation details. 
Section~\ref{sec:Results} shows our experimental results and main findings. 
\gao{Section~\ref{sec:Discussion} discusses the impact of different large language models, the influence of different data splits, the influence of similarity values on the assessment accuracy, the potential impact of the data leakage issue, and potential threats to our study.}
Section~\ref{sec:Related Work} summarizes related work and emphasizes our novelty. 
Section~\ref{sec:Conclusion} summarizes our study and discusses potential future directions.

\section{Background}
\label{sec:Background}

In this section, we first introduce the background of software vulnerability assessment.
Then we introduce the large language model and in-context learning.
Finally, we analyze the research challenges of applying LLMs to the SVA task.

\subsection{Software Vulnerability Assessment}

Software vulnerability assessment involves rating the severity of vulnerabilities based on their characteristics, which means calculating a risk score for each vulnerability according to scoring rules and determining its severity based on that score~\cite{le2022survey,humayun2022security}. SVA is critical because it helps organizations prioritize their efforts in addressing security risks. Identifying and accurately assessing the severity of vulnerabilities enables efficient allocation of resources to mitigate the most critical threats, thereby protecting systems and data from potential breaches. Inaccurate assessment could result in overlooking high-risk vulnerabilities or over-prioritizing low-risk ones, leading to inefficient use of security resources.

Assessing vulnerabilities manually is challenging due to the huge volume of vulnerabilities that modern software systems can contain. Manual assessment requires significant expertise and time, which can lead to inconsistencies and errors. Developers might also struggle with maintaining up-to-date knowledge of the ever-evolving landscape of security threats. The complexity and variability of software environments further complicate the manual assessment, making automation a valuable approach.

\begin{figure}[h!] 
    \centering 
    \includegraphics[scale=0.53]{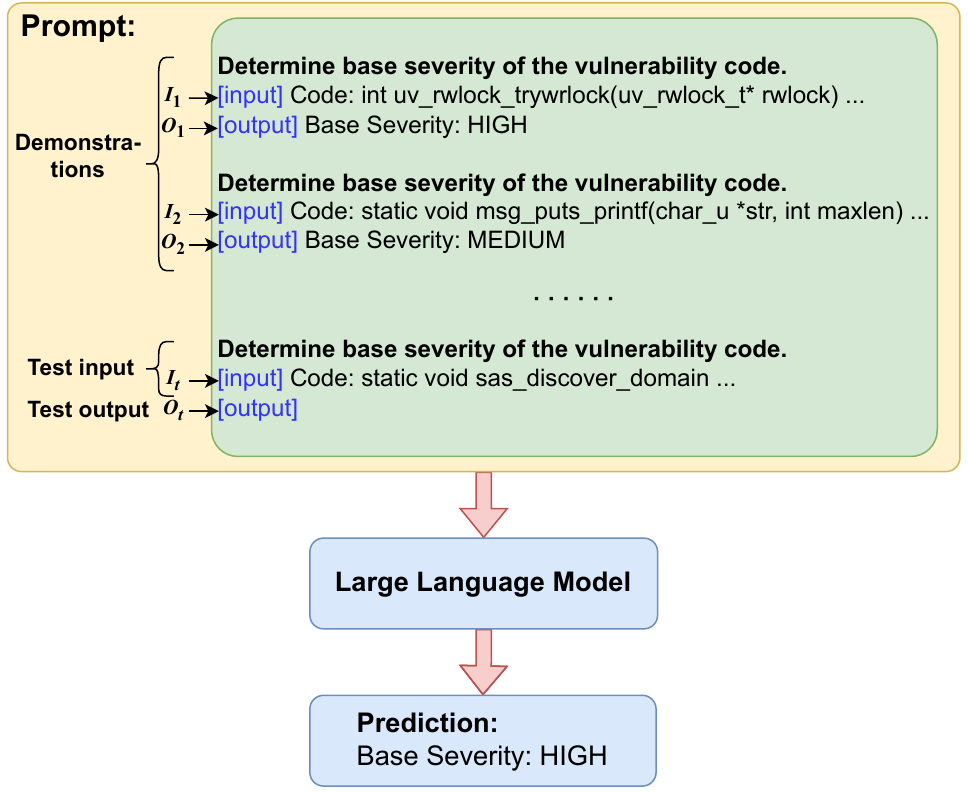} 
    \caption{An example of using in-context learning for software vulnerability assessment} 
    \label{Fig:ICL} 
\end{figure}

The Common Vulnerability Scoring System (CVSS)~\cite{cvss} is a standardized approach for measuring the severity of software vulnerabilities. It assesses vulnerabilities in terms of multiple dimensions (e.g., impact, exploitability) 
and provides a scoring system that offers organizations and researchers a standard for software vulnerability assessment. CVSS has been updated multiple times, with CVSS v3 being the most widely used version. 
In previous SVA studies, researchers mainly used CVSS v2~\cite{le2019automated,le2021deepcva} for SVA. However, compared to CVSS v2, CVSS v3 provides a more comprehensive, flexible, and accurate framework for assessing and prioritizing vulnerabilities, such as improved granularity, better exploitability metrics, and enhanced impact metrics. Therefore, we consider CVSS v3 in our constructed SVA dataset. 
In the realm of software vulnerability assessment, reliable and comprehensive data sources are crucial for both research and practical applications. One such valuable resource is the National Vulnerability Database (NVD)~\cite{le2022use,nvd}, a public database maintained by the National Institute of Standards and Technology (NIST). NVD serves as a critical repository, offering detailed information on software vulnerabilities, including descriptions, affected products and versions, patch information, and related reference links. In our research, we primarily use source code and vulnerability description information from NVD.

\subsection{Large Language Model and In-Context Learning}

Large Language Models (LLMs) are natural language processing models trained on large-scale text data, learning rich language knowledge and contextual relationships~\cite{kasneci2023chatgpt,chang2024survey}. 
In recent years, LLMs have achieved significant performance in various software engineering tasks, such as code generation~\cite{vaithilingam2022expectation,liu2024your}, code completion, source code summarization~\cite{vaithilingam2022expectation,geng2024large}, source code repair~\cite{tang2024code}, and vulnerability detection~\cite{purba2023software,zhou2024large}. 
However, to the best of our knowledge, LLMs have not yet been used for software vulnerability assessment.

In-context learning (ICL) is an approach that allows models to learn and handle tasks by providing demonstration examples without explicitly fine-tuning model parameters~\cite{dong2022survey,min2022rethinking}. The high flexibility, efficiency, and generality of ICL have attracted many researchers and can be extended to specific software engineering tasks.
Figure~\ref{Fig:ICL} illustrates an example of using in-context learning for software vulnerability assessment. In detail, we can use $I_i$, $O_i$, $I_i^{*}$, $O_i^{*}$, and $N$ to represent the initial input, initial output, reconstructed input, reconstructed output, and the number of demonstration examples, respectively. Here, the reconstructed input represents the initial input after optimization through the prompt template, and the reconstructed output is the corresponding output from the large language model. In this way, we can represent the initial set of $N$ demonstration examples as a collection $\{(I_1, O_1),(I_2, O_2),...,(I_N, O_N)\}$. Subsequently, we can fill these initial demonstration examples into a carefully designed prompt template, resulting in reconstructed demonstration examples $\{(I_1^{*}, O_1^{*}),(I_2^{*}, O_2^{*}),...,(I_N^{*}, O_N^{*})\}$. Typically, the number $N$ is fewer than 50 and is far less than the number of training examples required for fine-tuning methods~\cite{gao2023makes,wang2021codet5}. Therefore, in-context learning is also known as few-shot learning. Specifically, when $N$ is 0, it is referred to as zero-shot learning. We can represent the reconstructed demonstration examples as $D=I_{1}^{*}\parallel O_{1}^{*}\parallel I_{2}^{*}\parallel O_{2}^{*}\parallel...\parallel I_{N}^{*}\parallel O_{N}^{*}$ and further concatenate the test example $I_{t}^{*}$ with the demonstration examples to obtain the LLM's input prompt $P=D\parallel I_{t}^{*}$, where $\parallel$ denotes the literal concatenation operation. Finally, the LLM can use the input prompt $P$ to predict the test example's label $O_{t}$.

\subsection{Research Challenges}
\label{sec:motivation}

However, directly applying ICL to SVA presents several challenges. 
First, it is necessary to select high-quality demonstrations. Previous studies~\cite{le2022use,le2019automated} typically consider either the source code or the vulnerability description. 
However, in practice, considering both types of information simultaneously could be more effective, necessitating the design of efficient information fusion methods. The rationale evaluation of this information fusion setting can be found in Section~\ref{sec:resultRQ2}.
Additionally, for different modalities, it is crucial to develop corresponding similarity evaluation methods for source code and vulnerability description, respectively. 
Finally, it is important to analyze the impact of the number of demonstration examples and different ordering strategies for ICL on the performance of the LLM-based SVA approach.

\section{Approach}
\label{sec:Framework}

\begin{figure*}
	\centering
	\includegraphics[width=0.95\textwidth]{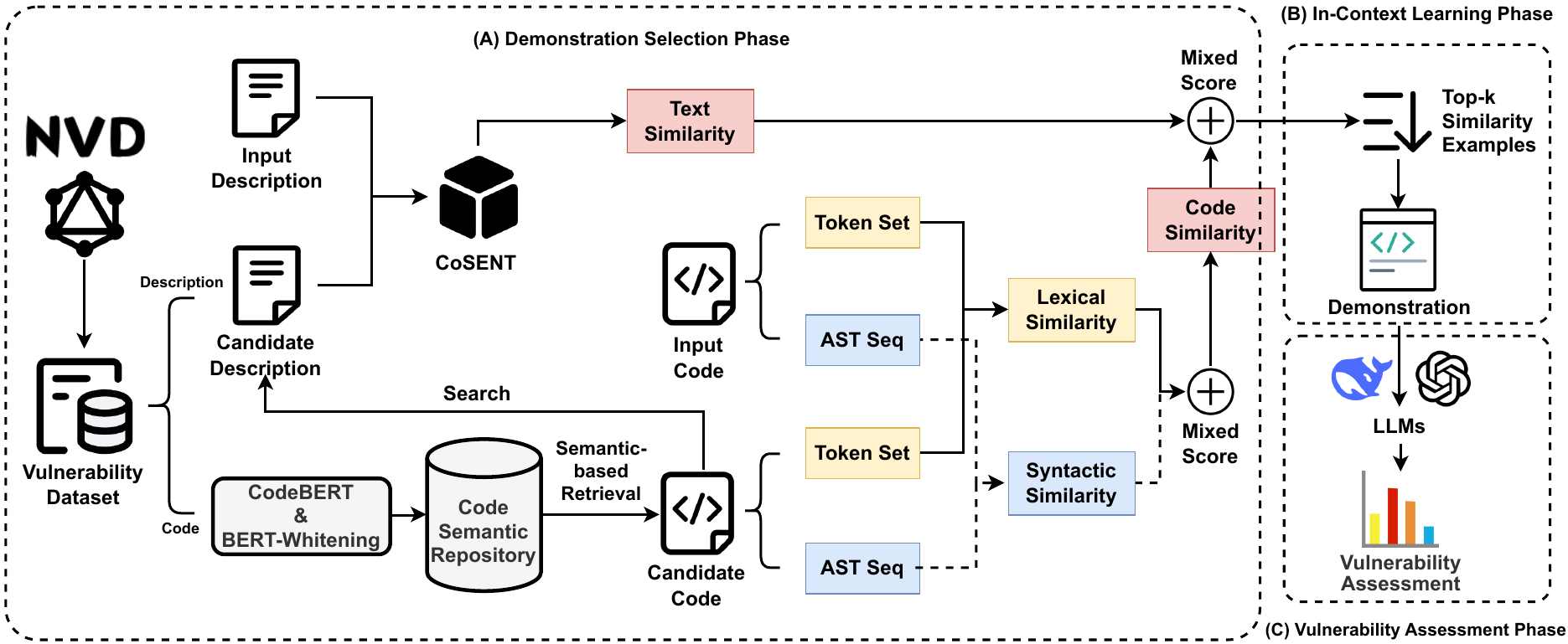}
	\caption{Framework of our proposed approach {\tool}}
	\label{fig:methodology}
\end{figure*}

Figure~\ref{fig:methodology} shows the framework of our proposed approach {\tool}. 
The framework of {\tool} mainly contains three stages. 
Specifically, in the \textbf{demonstration selection phase}, {\tool} selects the top-\emph{k} vulnerabilities as the demonstration examples from the historical repository by using information fusion to simultaneously consider both the source code and the vulnerability description. In the \textbf{in-context learning phase}, {\tool} inputs the retrieved top-$k$ vulnerabilities with their source code and vulnerability descriptions into our carefully designed prompting templates. 
Finally, in the \textbf{vulnerability assessment phase}, the input consists of the source code and its corresponding vulnerability description of the target vulnerability, and the output is the corresponding vulnerability severity level generated by the LLM.
In the remaining part of this section, we show the details of these three stages.

\subsection{Demonstration Selection Phase}

Recent research~\cite{gao2023makes} on in-context learning for code intelligence tasks (such as source code summarization, bug fixing, and program synthesis) has shown that the performance of ICL depends largely on the quality of demonstration examples. Therefore, carefully designed strategies for selecting demonstration examples from
the historical repository can effectively improve the performance of LLM in specific domain tasks~\cite{gao2023makes}. For the SVA task, to retrieve high-quality demonstration examples, we need to measure the similarity between different vulnerabilities. Some previous SVA studies only consider the vulnerability description, such as the multi-task learning method proposed by Babalau et al.~\cite{babalau2021severity}. Other previous SVA studies only consider the source code, such as the deep learning model DeepCVA proposed by Le et al.~\cite{le2021deepcva}.
However, we conjecture that considering only one type of vulnerability information is not comprehensive enough. Therefore, when calculating similarity, it is necessary to consider both types of vulnerability information by leveraging information fusion. Specifically, given a target vulnerability and the vulnerability from the historical repository, we first calculate the similarity based on the source code and then calculate the similarity based on the vulnerability description. Finally, we fuse the similarities of these two modalities.

Since how to accurately calculate code similarity is still an open problem for semantic-based software analysis, in this study, we consider an information retrieval-based method proposed by our previous study~\cite{yang2022ccgir} for calculating code similarity, which was successfully used in the LLM-based smart contract code comment generation task~\cite{zhao2024automatic}. This method simultaneously considers the semantic information, lexical information, and syntactic information of code snippets. Since {\tool} further considers vulnerability description for demonstration selection, we design the demonstration selection strategy by modifying our previously proposed information retrieval-based method, which can further consider the text similarity of vulnerability descriptions.

Specifically, our demonstration selection strategy can be divided into three parts: (1) \textbf{Semantic-based retrieval part:} We extract semantic information from vulnerable source code using CodeBERT~\cite{feng2020codebert} and BERT-whitening~\cite{su2021whitening}. Then, we retrieve the most similar source code from a historical dataset as candidates. (2) \textbf{Syntax and Lexicon-based retrieval part:} The first part only considers the semantic information of vulnerability codes, so in the second part, we compute the lexical and syntactic similarity of these candidate snippets, then combine the similarity of these two parts as the code similarity between the target vulnerability code snippet and the candidate vulnerability code snippet. (3) \textbf{Text-based retrieval part:} We retrieve candidate source code from the historical dataset obtained in the first part. In our dataset, each vulnerability code snippet corresponds to a vulnerability description. Therefore, we can find the corresponding vulnerability description based on the candidate source code. Then, we use the CoSENT model implemented based on Text2vec~\cite{Xu_Similarities_Compute_similarity} to calculate the text similarity between the target vulnerability description and the candidate vulnerability description. Finally, we combine the code similarity and text similarity to obtain the top-\emph{k} similar vulnerabilities, determining the final demonstration example. In the rest of this subsection, we provide detailed information for these three parts.

\subsubsection{Semantic-based retrieval part}

This part involves splitting the vulnerability code snippets in the historical dataset according to the CamelCase naming convention, then inputting the resulting sequences $\{X_i\}_{i=1}^N$ into CodeBERT~\cite{zhao2024automatic,liu2023automated} to obtain semantic vectors $X_i\in\mathbb{R}^D$, where $N$ represents the number of source code in the historical dataset, and $D$ represents the hidden dimension. Subsequently, BERT-whitening is used to improve the isotropy of the representations and reduce the dimension from $D$ to $d$, resulting in $\left\{\tilde{X}_{i}\right\}_{i=1}^{N}$. By reducing redundancy, shortening training time, and handling noisy data more effectively, using BERT-whitening can improve the quality of embeddings generated by CodeBERT ~\cite{yin2022efficient,jiang2022promptbert}. Finally, the semantic similarity between the code embeddings $\tilde{X}_{a}$ and $\tilde{X}_{b}$ is calculated using the L2 distance, with the following formula.
\begin{equation}
\mathrm{SemSim}\left(\tilde{X}_{a},\tilde{X}_{b}\right)=\sum_{i=1}^{d}\left(\tilde{X}_{a}[i]-\tilde{X}_{b}[i]\right)^{2}
\end{equation}

\subsubsection{Syntax and Lexicon-based retrieval part}

In the first part, we retrieve the top-\emph{n} most similar candidate source code from the historical dataset based on semantic similarity. In this part, we further consider the syntactic and lexical information of the vulnerable source code. Hence, we compute the syntactic and lexical similarity between the target vulnerability code and the vulnerability codes in the historical dataset. Specifically, we first employ the tree-sitter library\footnote{\url{https://github.com/tree-sitter/tree-sitter}} to generate the AST (Abstract Syntax Tree) sequences of the source code. Considering the lengthy nature of our vulnerability codes, we reference the method  SimSBT~\cite{yang2021comformer} to optimize the representation of the AST sequences. Lastly, we utilize code tokens to calculate lexical similarity and employ the AST sequences to compute syntactic similarity. Finally, combining these two similarity values to derive the final source code similarity. This approach comprehensively considers various aspects of the code, facilitating the retrieval of source code most similar to the target vulnerability code. Notice that the reason for prioritizing the calculation of semantic similarity before the calculation of syntactic and lexical similarity is its significant improvement in retrieval quality shown in previous studies~\cite{yang2022ccgir,siow2022learning}.

For two source code snippets $A$ and $B$, along with their corresponding AST sequences $\tilde{A}$ and $\tilde{B}$, we can calculate the syntactic similarity using the following formula.
\begin{equation}
\text{SynSim}(A, B)=\frac{\mathrm{sum}(\mathrm{len}(\tilde{A}),\mathrm{len}(\tilde{B}))-\mathrm{lev}}{\mathrm{sum}(\mathrm{len}(\tilde{A}),\mathrm{len}(\tilde{B}))}
\end{equation}
where $lev$ represents the Levenshtein distance~\cite{yujian2007normalized} between $\tilde{A}$ and $\tilde{B}$.

Concerning lexical similarity, we treat source code as a set of tokens. For two token sets $\mathrm{set}_A$ and $\mathrm{set}_B$, we determine lexical similarity using the Jaccard similarity~\cite{bag2019efficient}.
\begin{equation}
\text{LexSim}(A, B)=\frac{\mid\text{set}_A\cap\text{set}_B\mid}{\mid\text{set}_A\cup\text{set}_B\mid}
\end{equation}

After calculating syntactic similarity and lexical similarity, we can fuse these two similarity values by the following formula and use it as the code similarity between the target vulnerability code snippet and the candidate source code.
\begin{equation}
\begin{aligned}\mathrm{CodeSim}(A, B)&=\lambda\times\mathrm{SynSim}(A, B)\\&+(1-\lambda)\times\mathrm{LexSim}(A, B)\end{aligned}\label{eq4}
\end{equation}
where $\lambda$ is a parameter, which is used to adjust the weight between different similarity values. 
In our previous studies~\cite{yang2022ccgir,zhao2024automatic}, we show the effectiveness of the setting rationality by designing a set of ablation studies when compared with various variants of this information retrieval-based method.

\subsubsection{Text-based retrieval part}

In the first part, we can retrieve the top-$n$ most similar source code snippets in the historical dataset.
For each retrieved vulnerability, we can further consider the corresponding vulnerability description and calculate the text similarity between them.
In our study. We utilize the CoSENT model provided by the text2vec-based library similarities\footnote{\url{https://github.com/shibing624/similarities}}.
This model can compute the similarity between vulnerability descriptions and is referred to as $TextSim$. We assume that for the vulnerability $V_a$, the vulnerability description of the source code $A$ is denoted as $D_a$. For the vulnerability $V_b$, the vulnerability description of the source code $B$ is denoted as $D_b$. Then we can compute the similarity of different vulnerabilities using the following formula, which can further select the top-\emph{k} most similar vulnerabilities in the historical datasets by considering both source code and their respective vulnerability description.
\begin{equation}
\begin{aligned}\mathrm{Sim}(V_a,V_b)&=\phi\times\mathrm{CodeSim}(A,B)\\&+(1-\phi)\times\mathrm{TextSim}(D_a,D_b)\end{aligned}\label{eq5}
\end{equation}
where $\phi$ is a parameter used to adjust the weight between source code similarity and vulnerability description similarity.

\subsection{In-context Learning Phase}
\label{sec:3.2}

\begin{figure}[h!] 
    \centering 
    \includegraphics[scale=0.53]{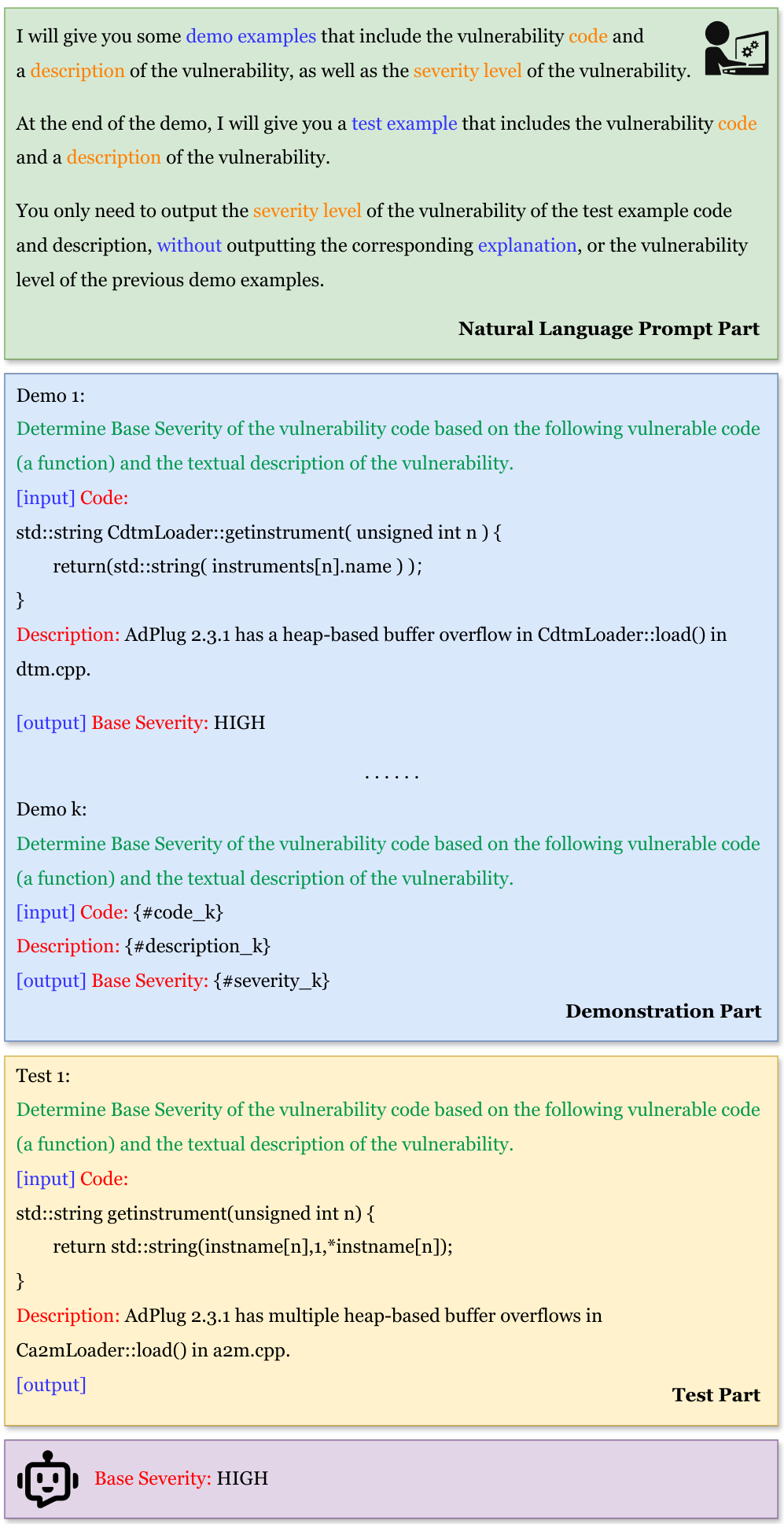} 
    \caption{The prompt template designed by our proposed approach {\tool}.} 
    \label{Fig:demonstration} 
\end{figure}

After retrieving high-quality demonstrations, we move on to designing the prompt template. The design of the prompt template is a crucial part of ICL, as high-quality templates can guide the large language model to output high-quality content, ensuring the accuracy and reliability of the information. 
Since we are the first to apply ICL to the SVA task, we mainly refer to the prompt templates designed in previous source code summarization studies~\cite{gao2023makes,zhao2024automatic}. Our designed prompt template is shown in Figure~\ref{Fig:demonstration}. 
Specifically, our prompt template consists of three parts. 
The first part is the \textbf{natural language prompt}, where we succinctly instruct the Large Language Model (LLM) to assess the severity of the target vulnerability based on the provided source code and vulnerability descriptions. To avoid LLM generating irrelevant redundant content, we emphasize instructing LLM to output only the base severity of the target vulnerability without additional explanations. 
In the \textbf{demonstration part}, we combine demonstration examples of source code and their corresponding vulnerability descriptions following a specific structure. 
First, we inform LLM to evaluate the severity of the vulnerability based on the source code and vulnerability description. 
Next, we use ``Demo $i$:" as the marker for the beginning of the $i$-th demonstration example. 
Then, we mark the start of input with ``[Input]:" to indicate the user's input. Within the input section, we use ``Code:" as the prompt for the source code and ``Description:" for the vulnerability description. 
Furthermore, we use ``[Output]:" to signify the start of the output section, which indicates the ground truth of this vulnerability. 
This organizational structure helps LLM better comprehend the input prompts and generate suitable responses~\cite{gao2023makes}. 
In the \textbf{test part}, we follow a similar prompt format as the demonstration part, with the distinction of changing ``Demo $i$:" to ``Test $i$:". Additionally, after ``[Output]:", no content is provided to clarify the specific output LLM needs to generate.

\subsection{Vulnerability Assessment Phase}

For the prompt template designed in Section~\ref{sec:3.2}, we can directly fill in the relevant content (such as the demonstration examples and the target vulnerability) into the template to generate the concrete prompt for LLM. Then, we can use the generated concrete prompt as input and call the API interface of the specific LLM. Finally, the API can output the severity level of the target vulnerability.

\section{Experimental Setup}
\label{Experimental Setup}

In this section, we first introduce our research questions and their design motivation. 
Then, we provide detailed information about the experimental subject, baselines, performance measures, and implementation details.

\subsection{Research Questions}

To show the competitiveness of {\tool} and the rationale of the component settings in {\tool}, we design the following five research questions (RQs) in our empirical study.

\textbf{RQ1: How does {\tool} perform in software vulnerability assessment compared to state-of-the-art baselines?} 

\textbf{Motivation.} 
In RQ1, we aim to analyze whether {\tool} can outperform state-of-the-art software vulnerability assessment baselines. 
Since our approach considers both source code and vulnerability descriptions. Therefore, when selecting SVA baselines, we consider both source code-based baselines and vulnerability description-based baselines. For example, for source code-based baselines, we consider baselines, such as $\text{Func}_{RF}$ and $\text{Func}_{LGBM}$~\cite{le2022use}. For vulnerability description-based baselines, we consider baselines, such as $\text{CWM}_{NB}$, $\text{CWM}_{SVM}$, and $\text{CWM}_{LR}$~\cite{le2019automated}. 
\gao{In addition, we also consider research baselines, such as Cvss-bert~\cite{shahid2021cvss}, SPSV~\cite{babalau2021severity}, SVA-PT~\cite{xue2025towards}, and MTLM~\cite{du2024vulnerability}.}

Since we model the SVA problem as a multi-class classification problem, we use widely adopted performance measures, such as Accuracy, Macro F1 score, and Matthews correlation coefficient, to evaluate the performance of {\tool} and SVA baselines.

\textbf{RQ2: Whether considering both source code and vulnerability description with information fusion can improve the performance of {\tool}?}

\textbf{Motivation.} 

Previous software vulnerability assessment studies mainly focused on either source code or vulnerability descriptions~\cite{le2022survey,le2022use,le2019automated,le2021deepcva}. In our study, we conjecture that considering both types of modalities simultaneously can help to select higher-quality demonstrations. To validate this conjecture, we design an ablation study in RQ2 and try to find the optimal fusion ratio of these different modalities for {\tool} in our study.

\textbf{RQ3: How does the number of demonstration examples in the prompt affect the performance of {\tool}?} 

\textbf{Motivation.} 
There are two reasons for designing RQ3. One reason is that existing work~\cite{zhao2024automatic} shows that the number of demonstrations affects the effectiveness of ICL. The other reason is the context token limitation in LLM~\cite{xue2024repeat,he2024talk}. If the input exceeds the token limitation, it will be truncated. Therefore, we want to investigate the impact of the number of demonstration examples on the performance of {\tool}.

\textbf{RQ4: How do different demonstration ordering strategies affect the performance of {\tool}?} 

\textbf{Motivation.} 
In RQ3, we mainly focus on analyzing the performance impact of the demonstration number on {\tool}. 
However, previous work~\cite{gao2023makes} indicates that the ordering strategy can also significantly affect performance. 
Therefore, in RQ 4, we aim to identify the optimal demonstration ordering strategy for {\tool} in our study.

\textbf{RQ5: How does the fusion ratio of syntactic similarity and lexical similarity for source code similarity affect the performance of {\tool}?} 

\textbf{Motivation.} 
Although the effectiveness of customization of our proposed information retrieval-based method has been validated in the smart contract code comment generation task~\cite{yang2022ccgir,zhao2024automatic}, whether this customization is effective in the SVA task still needs investigation. Therefore, in RQ5, we mainly investigate the performance influence of source code similarity setting (i.e., the fusion ratio of syntactic similarity and lexical similarity ) on {\tool}.

\subsection{Experimental Subject}

Our dataset is constructed using the raw data of MegaVul shared by Ni et al.~\cite{MegaVul}. The MegaVul dataset\footnote{\url{https://github.com/Icyrockton/MegaVul}}  was last updated on April 14, 2024. 
This dataset was crawled from the Common Vulnerabilities and Exposures (CVE) database~\cite{cve} and consists of 17,975 vulnerable data entries in C/C++ code. It covers 176 different types of vulnerabilities from 1,062 open-source projects, spanning from 2006 to 2024. It not only ensures code integrity through advanced tools but also provides various information dimensions (such as function signatures, abstract functions, and code changes). This dataset supports multiple software security tasks, including software vulnerability detection and fixing.

The severity of vulnerabilities is typically determined according to the Common Vulnerability Scoring System (CVSS) standard, a free and open industry standard. In this study, we consider the CVSS v3 standard. Under this standard, the severity of vulnerabilities is divided into four levels\footnote{\url{https://www.first.org/cvss/v3.1/specification-document}} according to the score values: Critical (9.0$\sim$10.0), High (7.0$\sim$8.9), Medium (4.0$\sim$6.9), and Low (0.1$\sim$3.9).
Compared with CVSS v2 used by previous SVA studies~\cite{le2022use,le2019automated,le2021deepcva}, CVSS v3 provides a more refined assessment of vulnerability severity and risk. It includes additional sub-scores and metrics, as well as added temporal and environmental metrics, which more accurately reflect the actual impact of vulnerabilities. 

After manual analysis, we found that the MegaVul dataset did not fully capture the severity ratings of all vulnerability scores according to the CVSS v3 standard. Therefore, we re-crawled the severity ratings and vulnerability scores of all vulnerabilities according to the CVSS v3 standard, along with their descriptions. After this step, we obtained 14,235 entries of vulnerability data following the CVSS v3 standard, and 3,740 entries without CVSS v3 information were removed. Then, we utilized the AST sequences of vulnerability code for syntactic similarity calculations. Specifically, we used the tree-sitter library\footnote{\url{https://tree-sitter.github.io/tree-sitter/}} to extract the AST sequence from the source code. We removed 2,164 entries that could not be successfully processed by the tree-sitter library due to syntax errors in the code or incomplete grammar rules. After this step, we ultimately gathered 12,071 vulnerability entries in our constructed dataset.
Though CVSS V4 has introduced new concepts and improvements recently. However, vulnerability data labeled with CVSS V4 scores remain very scarce in practice. The majority of scores in our constructed vulnerability datasets are still based on CVSS V2 and CVSS V3, resulting in a limited sample size for CVSS V4, which might affect the generalizability and reliability of the research findings. Therefore, in our study, we primarily base our analysis on CVSS V3.

Following previous studies~\cite{le2019automated,le2021deepcva}, we split our constructed dataset into the training set (80\%), the validation set (10\%), and the test set (10\%) by stratified sampling, ensuring that the proportions of the four vulnerability severity levels are consistent across these three sets. 
The statistical information of our experiment subject is shown in Table~\ref{tab1}. 
In this table, we first summarize the number of entries with different severity levels in the different datasets. For example, there are 1,169 vulnerabilities of critical severity in the training set, 146 in the validation set, and 147 in the test set. Next, we analyze the mean and median token counts in source code and vulnerability descriptions across the different datasets. For instance, the mean number of tokens in the source code is 616 in the training set, 632 in the validation set, and 577 in the test set.

\begin{table}[htb]

\centering
\caption{Statistical information of our experimental subject.}
\resizebox{8.6cm}{!}
{
\begin{tabular}{lccc}
\toprule
\textbf{Statistic} & \textbf{\makecell[c]{Train}} & \textbf{\makecell[c]{Validation}} & \textbf{\makecell[c]{Test}} \\
\midrule
Number & 9,656 & 1,207 & 1,208 \\
Number of Critical severity & 1,169 & 146 & 147 \\
Number of High severity & 4,454 & 577 & 577 \\
Number of Medium severity & 3,795 & 474 & 475 \\
Number of Low severity & 238 & 30 & 29 \\\midrule
Average tokens in codes & 616 & 632 & 577 \\
Average tokens in descriptions & 71 & 69 & 72 \\
Median tokens in codes & 334 & 361 & 331 \\
Median tokens in descriptions & 57 & 57 & 59 \\
\bottomrule
\end{tabular}
}
\label{tab1} 
\end{table}

\subsection{Baselines}

\gao{To evaluate the effectiveness of our proposed approach {\tool}, we compare it with two baselines based on vulnerability source code (i.e., $\text{Func}_{RF}$ and $\text{Func}_{LGBM}$~\cite{le2022use}), five baselines based on vulnerability descriptions (i.e., $\text{CWM}_{NB}$, $\text{CWM}_{SVM}$, $\text{CWM}_{LR}$~\cite{le2019automated}, Cvss-bert~\cite{shahid2021cvss} and SPSV~\cite{babalau2021severity}) and two baselines based on both vulnerability source code and descriptions (i.e., SVA-PT~\cite{xue2025towards} and MTLM~\cite{du2024vulnerability}). In the rest of this subsection, we introduce the characteristics of these baselines as follows.}

Le et al.~\cite{le2022use} proposed a source code-based vulnerability assessment approach, aimed at conducting evaluations at the function level. Their research considers the contextual information of the vulnerable source code, including lines of code that are directly or indirectly related to the vulnerability, to enhance the performance of the vulnerability assessment model. Depending on the chosen classifier, the approach can be divided into two separate approaches as follows.

\begin{itemize}
    \item \textbf{$\text{Func}_{RF}$}. This approach uses the Random Forest (RF) as the classifier. RF~\cite{ho1995random} is an ensemble learning approach that performs classification by building multiple decision trees and uses the concept of ensemble learning to improve overall accuracy and stability. The main hyperparameters of RF are the number of trees, maximum depth, and maximum number of leaves. 
    
    \item \textbf{$\text{Func}_{LGBM}$}. This approach uses the Light Gradient Boosting Machine (LGBM) as the classifier.  LGBM~\cite{ke2017lightgbm} is a fast, scalable, and high-performance gradient-boosting decision tree algorithm that belongs to the Gradient Boosting Framework. Its main advantages lie in efficiency, scalability, and accuracy, enabling it to quickly train high-performance models on large datasets. The main hyperparameters are similar to RF.
\end{itemize}

In addition to the source code-based approach, Le et al.~\cite{le2019automated} proposed a systematic approach based on vulnerability descriptions that integrates both character and word features. They also devised a time-based cross-validation technique for determining the optimal model for each vulnerability characteristic. They created natural language processing (NLP) configurations using different values of n-grams in combination with either tf (term frequency) or tf-idf (term frequency-inverse document frequency) weighting. Similar to the first baseline, this approach can be divided into three approaches depending on the chosen classifier.

\begin{itemize}
    \item \textbf{$\text{CWM}_{NB}$}.  This approach uses the Na{\"i}ve Bayes (NB) as the classifier. NB~\cite{russell2016artificial} is a simple probabilistic classifier based on Bayes' theorem and the assumption of feature conditional independence. NB excels in efficiency with large amounts of data, fast training, easy implementation, and outstanding performance in classification tasks. Following the study of Le et al.~\cite{le2019automated}, no hyperparameters were optimized during the validation process for NB.

    \item \textbf{$\text{CWM}_{SVM}$}. This approach uses the Support Vector Machine (SVM) as the classifier. SVM~\cite{cortes1995support} is a supervised learning algorithm that distinguishes different classes by finding an optimal hyperplane in the feature space. For multi-class problems, we use the one-vs-rest scheme to convert them into multiple binary classification problems.

    \item \textbf{$\text{CWM}_{LR}$}. This approach uses Logistic Regression (LR) as the classifier. LR~\cite{walker1967estimation} is a statistical learning approach used to solve binary classification problems, and it can be extended to multi-class problems using the one-vs-rest scheme. Note that the optimal hyperparameter values for these three classifiers may vary when using different NLP representations such as tf and tf-idf.

\end{itemize}

\gao{For these baselines based on machine learning methods, we directly used their shared code and performed hyperparameter optimization.} The hyperparameter names, candidate values, and optimal values for different baselines are shown in Table~\ref{tab2}.

\gao{Shahid et al.~\cite{shahid2021cvss} proposed a SVA approach based on natural language processing. They trained multiple BERT classifiers, one for each CVSS vector metric, to determine the CVSS vector and severity score from vulnerability textual descriptions. A gradient-based input saliency method was used for explainability. For convenience, we refer to this approach as Cvss-bert.}

\gao{Babalau et al.~\cite{babalau2021severity}  proposed a sVA approach based on vulnerability text descriptions. They used a Multi-Task Learning architecture with a pre-trained BERT model to compute vector-space representations of words in vulnerability descriptions, aiming to predict the severity score and other metrics of a vulnerability using only its text description available upon discovery. We refer to this approach as SPSV.}

\gao{Xue et al.~\cite{xue2025towards} applied prompt tuning and continual learning for SVA. They combined confidence-based replay and regularization, used source code and vulnerability descriptions to create hybrid prompts for tuning CodeT5. It outperformed baselines, and ablation studies verified its component effectiveness. We refer to this approach as SVA-PT.}

\gao{Du et al.~\cite{du2024vulnerability} proposed a vulnerability severity prediction approach MTLM. Specifically, they preprocessed bimodal data (i.e., vulnerability description and source code), used GraphCodeBert for feature extraction, a Bi-GRU with attention for further feature extraction, and a multi-task learning approach with hard parameter sharing.}

\newcolumntype{L}[1]{>{\raggedright\arraybackslash}p{#1}}
\newcolumntype{C}[1]{>{\centering\arraybackslash}p{#1}}
\newcolumntype{R}[1]{>{\raggedleft\arraybackslash}p{#1}}
\begin{table*}[htb]

\centering
\caption{\fontsize{10.5bp}{17bp}The range of values and the optimal value for baseline's hyperparameters.}
\resizebox{18.3cm}{!}
{
\begin{tabular}{{L{4cm}L{5cm}L{6cm}L{4cm}}}

\toprule
\textbf{Baseline} & \textbf{Hyperparameter Name} & \textbf{Candidate Values} & \textbf{Optimal Value} \\
\midrule
$\text{Func}_{RF}$ & \emph{No. of estimators} \newline
\emph{Max depth} \newline
\emph{Max. no. of leaf nodes} & 100, 200, 300, 400, 500 \newline 3, 5, 7, 9, unlimited \newline 100, 200, 300, unlimited & 200 \newline unlimited \newline 300\\

\hline
$\text{Func}_{LGBM}$ & \emph{No. of estimators} \newline
\emph{Max depth} \newline
\emph{Max. no. of leaf nodes} & 100, 200, 300, 400, 500 \newline 3, 5, 7, 9, unlimited \newline 100, 200, 300, unlimited & 200 \newline unlimited \newline 100 \\

\hline
$\text{CWM}_{NB}$ & None & None & None \\

\hline
$\text{CWM}_{SVM}$ & \emph{Kernel} \newline
\emph{Regularization coefficient} & linear \newline 0.01, 0.1, 1, 10, 100 & linear \newline 0.1 \\

\hline
$\text{CWM}_{LR}$ & \emph{Regularization coefficient} & 0.01, 0.1, 1, 10, 100 & 0.1 for tf \newline
10 for tf-idf \\

\bottomrule
\end{tabular}
}
\label{tab2} 
\end{table*}

\subsection{Performance Measures}
\label{sec:Performance Measure}

To assess the performance of our proposed approach {\tool}, we utilize commonly used performance measures in previous software vulnerability assessment studies~\cite{le2022use,le2021deepcva}, such as Accuracy, F1-score, and MCC (Matthews Correlation Coefficient). The latter two measures are particularly suitable for our constructed dataset, which has a class imbalanced problem~\cite{luque2019impact}, as shown in Table~\ref{tab1}. The details of these performance measures are as follows:

\begin{itemize}
    \item \textbf{Accuracy}. Accuracy represents the proportion of correctly classified samples among all samples, measuring the overall classification performance of the approach. The range of Accuracy values is from 0 to 1, where 0 indicates all samples are classified incorrectly, and 1 indicates all samples are classified correctly. A higher value signifies better performance.

    \item \textbf{F1-score}. The F1-score is the harmonic mean of precision and recall. To evaluate the performance of an approach across the entire dataset in multi-class tasks, we utilize the macro F1-score~\cite{spanos2018multi}. The range of F1-score values is from 0 to 1, with values closer to 1 indicating better performance, demonstrating a good balance between precision and recall.

    \item \textbf{MCC}. MCC is a crucial performance measure for multi-class tasks. MCC is particularly useful for handling class-imbalanced datasets as it considers the predictive performance of each class~\cite{luque2019impact}. Similar to the F1-score, we use the multi-class version of MCC~\cite{gorodkin2004comparing}. MCC values range from -1 to 1, with values closer to 1 indicating better performance, meaning the approach's predictions in the classification task are more accurate.
\end{itemize}

By using these three performance measures, we can comprehensively assess the performance of {\tool} and baselines in the software vulnerability assessment task, thereby providing a thorough performance analysis.

\subsection{LLM used by {\tool}}

In contrast to previous LLM-based software engineering studies~\cite{zhao2024automatic,ye2023comprehensive,anand2023gpt4all}, we did not use GPT-3.5 Turbo as the primary LLM for {\tool}. Instead, we employed DeepSeek-Chat (DeepSeek-V2)~\cite{deepseekv2} for several reasons.
Firstly, DeepSeek-V2, released in May 2024, is the latest version with capabilities close to GPT-4, surpassing GPT-3.5 Turbo comprehensively~\cite{deepseekv2}. Secondly, due to the extensive ablation studies in our research (such as RQ2 to RQ5), we chose DeepSeek-V2 to reduce experimental costs, as the price of DeepSeek-V2 is only about one-third of GPT-3.5 Turbo. Thirdly, the length of the API input content in LLMs is limited by the context window length. Following previous studies~\cite{gao2023makes}, we truncate content exceeding the token limitation. However, previous research also showed that truncation could lead to performance degradation~\cite{gao2023makes,dai2019transformer}. Given that our dataset's source code and vulnerability descriptions are relatively long (see Table~\ref{tab1}), we selected the DeepSeek-V2 LLM with a longer context window to mitigate performance decreases due to truncation. Specifically, the token limitation for the DeepSeek-V2 LLM is 32k tokens, while the GPT-3.5 Turbo LLM is only 16k tokens. Detailed comparison results of using different LLMs for {\tool} can be found in Section~\ref{sec:Discussion}. 
Regarding the cost of using large language models in this experiment, we acknowledge that different large language model products have varying pricing strategies, although they typically use the same token calculation method. As a reference, we are providing the total token consumption for our experiments. Based on backend records, all experiments in this study consumed approximately 108 million tokens.

\subsection{Implementation Details}
\label{sec:Implementation Details}

We constructed prompt templates using four demonstration examples, arranged in ascending order based on their similarity. To measure code similarity, we set the ratio of syntactic to lexical similarity at 60\% to 40\%. For information fusion, we set the ratio of source code similarity to vulnerability description similarity at 70\% to 30\%. The rationale for these experimental settings is detailed in our ablation study analysis.

Regarding the hyperparameters of the LLM, and following previous studies~\cite{gao2023makes,nashid2023retrieval,cheng2022binding}, we set the temperature parameter to 0. Both the ``frequency\_penalty" and ``presence\_penalty" are also set to 0. Specifically, setting the temperature to 0 allows the model to provide the most probable prediction, resulting in text generation with higher certainty. With a frequency penalty of 0, the model does not penalize repeated occurrences of words. Similarly, with a presence penalty of 0, the model does not emphasize the presence of specific words or phrases.

We conduct all experiments on a server equipped with a GeForce RTX 4090 GPU (24GB graphics memory). The server runs on the Windows 10 operating system.

\section{Experimental Results}
\label{sec:Results}

\subsection{RQ1: How does {\tool} perform in software vulnerability assessment compared to state-of-the-art baselines?}

\textbf{Approach:}

\gao{To demonstrate the effectiveness of our proposed approach {\tool} for the software vulnerability assessment task, we select $\text{Func}_{RF}$, $\text{Func}_{LGBM}$~\cite{le2022use}, $\text{CWM}_{NB}$, $\text{CWM}_{SVM}$, $\text{CWM}_{LR}$~\cite{le2019automated}, Cvss-bert~\cite{shahid2021cvss}, SPSV~\cite{babalau2021severity}, SVA-PT~\cite{xue2025towards} and MTLM~\cite{du2024vulnerability} as state-of-the-art SVA baselines.} 
For {\tool}, we consider the experimental settings shown in Section~\ref{sec:Implementation Details}. Performance evaluation of these approaches was conducted using the measures introduced in Section~\ref{sec:Performance Measure}.

\textbf{Results:} 
The performance comparison results between our proposed approach {\tool} and the baselines are presented in Table~\ref{tab:RQ1}. The optimal results for each performance measure are highlighted in bold, while the second-best results are underscored. The results indicate that our proposed approach {\tool} outperforms all baselines in terms of three performance measures, particularly excelling in the MCC measure. \gao{Specifically, {\tool} achieves 77.07\% for Accuracy, 67.89\% for F1-score, and 63.01\% for MCC. Compared to the other baseline approaches, {\tool} improves Accuracy by at least 2.53 pp (percentage point), F1-score by at least 3.04 pp, and MCC by at least 3.49 pp.}

\begin{table}[htb]

\centering
\caption{Comparison results between {\tool} and baselines in terms of three performance measures.}
\resizebox{8.6cm}{!}
{
\begin{tabular}{lccc}
\toprule
\textbf{Approach} & \makecell[c]{\textbf{Accuracy (\%)}} & \makecell[c]{\textbf{F1-score (\%)}} & \makecell[c]{\textbf{MCC (\%)}} \\
\midrule
$\text{Func}_{RF}$ & 62.97 & 43.69 & 37.25 \\
$\text{Func}_{LGBM}$ & 69.39 & 62.48 & 48.51 \\
$\text{CWM}_{NB}$ & 62.44 & 49.26 & 37.98 \\
$\text{CWM}_{SVM}$ & 68.16 & 58.53 & 46.74 \\
$\text{CWM}_{LR}$ & 69.32 & 62.55 & 49.42 \\
\gao{Cvss-bert} & 70.31 & 61.26 & 50.54 \\
\gao{SPSV} & 72.42 & 61.76 & 53.68 \\
\gao{SVA-PT} & 72.55 & 62.43 & 56.13 \\
\gao{MTLM} & \uline{74.54} & \uline{64.85} & \uline{59.52} \\
{\tool} & \textbf{77.07} & \textbf{67.89} & \textbf{63.01} \\
\bottomrule
\end{tabular}
}
\label{tab:RQ1} 
\end{table}

\gao{The cost of calculating the code similarity for the target vulnerability is an important factor in evaluating the effectiveness of {\tool}. Although similarity computation requires certain computational resources, the vector representations of vulnerability code snippets can be pre-computed offline, thereby significantly reducing the online computation burden. Specifically, during the preliminary preparation stage, we employ CodeBERT and BERT-whitening to process the vulnerability code snippets in the dataset, generating their semantic vectors and optimizing the vector representations. While this offline computation process does require an initial investment of computational resources, it is a one-time cost. During the subsequent vulnerability assessment, when calculating the similarity between a target vulnerability and historical vulnerabilities, we can directly leverage the pre-computed vector representations, substantially reducing the cost of real-time computation.}

\begin{tcolorbox}[width=\linewidth, title={}]
\textbf{Summary for RQ1:} The experimental results demonstrate that {\tool} outperforms baselines in software vulnerability assessment. \gao{Specifically, it shows improvements of 2.53 pp in Accuracy, 3.04 pp in F1-score, and 3.49 pp in MCC compared to the second-best baseline.}
\end{tcolorbox}

\subsection{RQ2: Whether considering both source code and vulnerability description with information fusion can improve the performance of {\tool}?}
\label{sec:resultRQ2}

\textbf{Approach:} In this RQ, we examine the impact of integrating information from different modalities on the performance of {\tool}. We combine code similarity for source code and text similarity for vulnerability descriptions using varying ratios based on Equation~\ref{eq5}, which assigns different weights to each modality. We explore 11 different ratios, such as 0\%:100\% (considering only vulnerability descriptions), 50\%:50\% (equal weighting of both modalities), and 100\%:0\% (considering only source code).

\textbf{Results:} Table~\ref{tab:RQ2} presents the performance of our proposed approach {\tool} under 11 different fusion ratios of code similarity and text similarity. Specifically, when the code similarity and text similarity ratios are set to 70\% and 30\%, respectively, our proposed approach {\tool} achieves the best performance in terms of the MCC measure. This indicates that source code information is more critical for assessing the base severity of vulnerabilities. However, incorporating vulnerability descriptions helps the LLM better understand the severity of the vulnerability and make more accurate assessments. When the text similarity ratio is set to 100\%, i.e., without considering source code, our proposed approach {\tool} performs poorly across all three performance measures. This suggests that examples selected using only vulnerability descriptions are of low quality, making it nearly impossible for the LLM to correctly assess the base severity of vulnerabilities. Conversely, when the code similarity ratio is set to 100\% (i.e., without considering vulnerability descriptions), the performance of {\tool} also decreases. This further confirms that integrating both vulnerability descriptions and source code improves the performance of {\tool} in the software vulnerability assessment task.

\begin{table}[htb]
\centering
\caption{Experimental results on the ratio of similarity between source code and vulnerability description in our proposed approach {\tool}. Notice that CodeSim represents the ratio of similarity in the source code, and TextSim represents the ratio of similarity in the vulnerability descriptions.}
\resizebox{8.6cm}{!}
{
\begin{tabular}{ccccc}
\toprule
\textbf{CodeSim} & \textbf{TextSim} & \makecell[c]{\textbf{Accuracy (\%)}} & \makecell[c]{\textbf{F1-score (\%)}} & \makecell[c]{\textbf{MCC (\%)}} \\
\midrule
100\% & 0\% & 74.25 & 63.56 & 58.14 \\
90\% & 10\% & 76.08 & 67.66 & 61.43 \\
80\% & 20\% & 75.91 & \textbf{68.34} & 61.24 \\
70\% & 30\% & \textbf{77.07} & 67.90 & \textbf{63.01} \\
60\% & 40\% & 75.75 & 67.40 & 60.93 \\
50\% & 50\% & 76.32 & \uline{68.16} & 61.82 \\
40\% & 60\% & 75.75 & 66.80 & 60.87 \\
30\% & 70\% & \uline{76.41} & 67.83 & \uline{62.04} \\
20\% & 80\% & 75.83 & 67.14 & 61.15 \\
10\% & 90\% & 75.99 & 66.98 & 61.52 \\
0\% & 100\% & 46.77 & 34.72 & 13.65 \\
\bottomrule
\end{tabular}
}
\label{tab:RQ2} 
\end{table}

\begin{figure}[h!] 
    \centering 
    \includegraphics[scale=0.53]{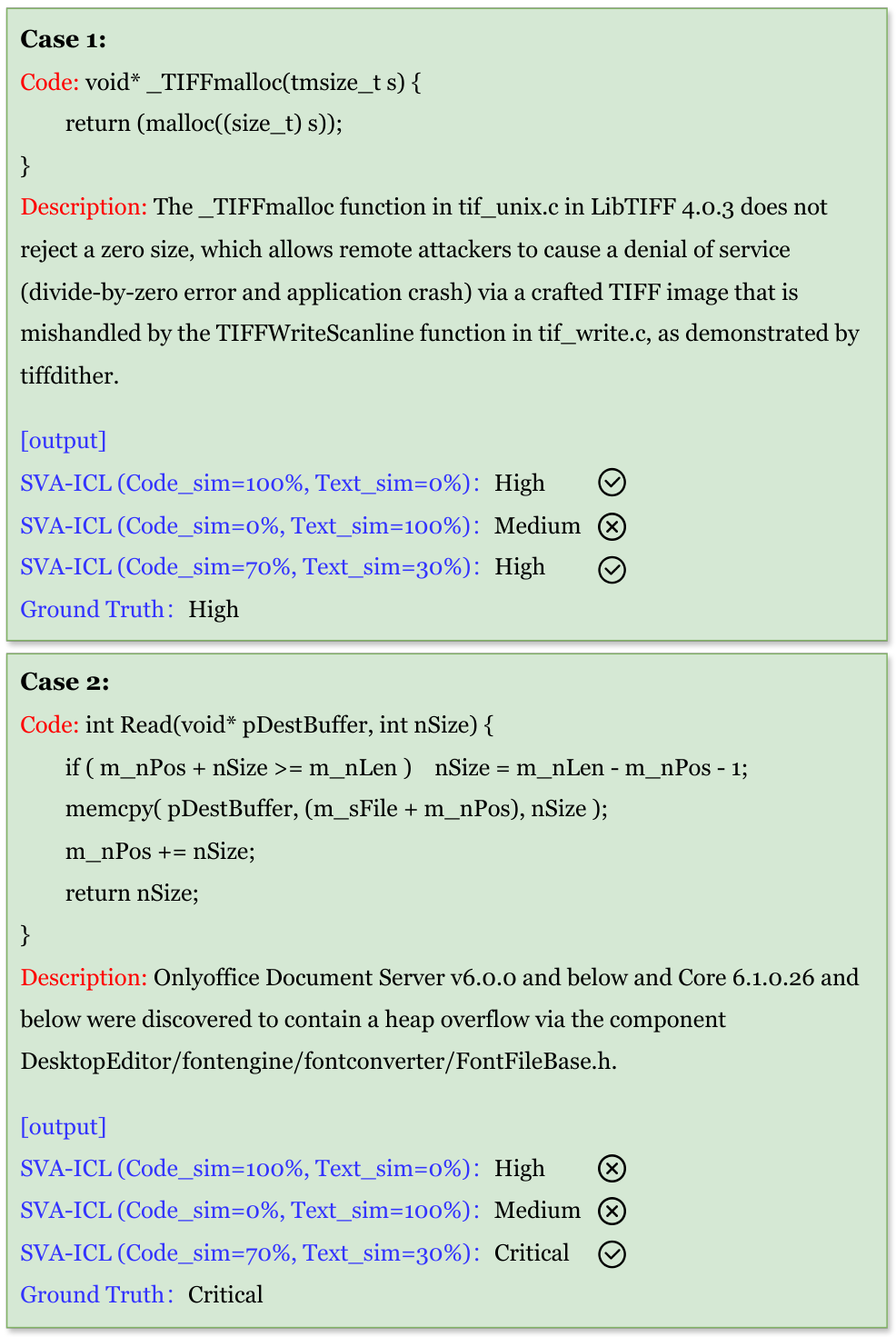} 
    \caption{Two cases of the base severity returned by our proposed approach {\tool} (which only considers source code, only considers vulnerability description, and considers both with the best ratio setting)} 
    \label{Fig:case} 
\end{figure}

In addition to quantitative analysis based on evaluation measures, we also perform qualitative analysis using two cases. The results of these two cases with different ratio settings are shown in Figure~\ref{Fig:case}. In Case 1, when the similarity of source code is not considered, {\tool} returns an incorrect assessment result (i.e., the ground-truth value is "High", while the assessment value is "Medium"). Further analysis reveals that this may be due to excessive redundancy in the vulnerability description, making it difficult for the LLM to accurately assess the severity level. However, when the source code is considered, our proposed approach {\tool} makes the correct assessment, demonstrating that incorporating both source code and vulnerability description information helps the LLM in software vulnerability assessment. In Case 2, {\tool} predicts incorrectly when considering only source code or only vulnerability descriptions, but makes the correct assessment when both bimodal information sources are considered. These two real-world cases further demonstrate that integrating code and description bimodal information and using a reasonable ratio setting helps improve the performance of {\tool}.

\begin{tcolorbox}[width=1.0\linewidth, title={}]
\textbf{Summary for RQ2:} Compared to considering a single modality, integrating two modalities (i.e., source code and vulnerability descriptions) achieves better performance. Moreover, a ratio setting of 70\%:30\% achieves the optimal performance of {\tool} in our study.
\end{tcolorbox}

\subsection{RQ3: How does the number of demonstration examples in the prompt affect the performance of {\tool}?}

\textbf{Approach:} 
 Inspired by previous work~\cite{zhao2024automatic}, we investigate the impact of the number of demonstration examples on the performance of our proposed approach {\tool}. Specifically, during the demonstration example selection phase, we select the top 10 candidate vulnerabilities most similar to the target vulnerability. We then construct ICL demonstration examples using 0, 1, 4, and 5 of the most similar vulnerabilities. According to the results of RQ2, we set the ratio of code similarity to text similarity at 70\% and 30\% in RQ3. Additionally, the length of source code and vulnerability descriptions in our dataset is relatively long, which is constrained by the context window length of the LLM. Therefore, we truncate the prompt template as described in Section~\ref{sec:Implementation Details}. We use the same prompt template for different numbers of demonstration examples as shown in Figure~\ref{Fig:demonstration} to ensure a fair comparison.

\textbf{Results:} 
We show the performance of {\tool} with different numbers of demonstration examples in Table~\ref{tab:RQ3}. As shown in the table, when using zero-shot learning (i.e., without demonstration example), {\tool} performs the worst in terms of all three performance measures, with an MCC of only 7.14\%, which is a 55.87\% performance decrease compared to the best experimental setting. This indicates that without considering demonstration examples, it is difficult to utilize the relevant knowledge in the LLM. Comparing the results of using one demonstration example (i.e., with one-shot) versus four demonstration examples, it is evident that the number of demonstration examples significantly enhances the performance of LLMs. However, when using five demonstration examples to construct the prompt template, the performance decreases compared to using four examples. Based on our manual analysis and the findings of previous studies~\cite{dai2019transformer,bulatov2022recurrent}, the performance decrease may be caused by the content truncation of the prompt. This truncation may remove crucial information from the source code and vulnerability descriptions, hindering the LLM's ability to accurately understand the target vulnerabilities and leading to a performance drop. In this RQ, we find that setting the number of demonstration examples to 4 achieves optimal performance. In a previous study on code generation task~\cite{gao2023makes}, Gao et al. also recommend setting the demonstration number to 4 can help to achieve optimal performance. Therefore, the findings of our study are consistent with theirs.

\begin{table}[htb]

\centering
\caption{The performance impact of different demonstration numbers for our proposed approach {\tool}.}
\resizebox{8.6cm}{!}
{
\begin{tabular}{lccc}
\toprule
\textbf{Setting} & \makecell[c]{\textbf{Accuracy (\%)}} & \makecell[c]{\textbf{F1-score (\%)}} & \makecell[c]{\textbf{MCC (\%)}} \\
\midrule
with zero-shot & 44.21 & 27.53 & 7.14 \\
with one-shot & 66.39 & 54.18 & 45.73 \\
with 4-shot & \textbf{77.07} & \textbf{67.89} & \textbf{63.01} \\
with 5-shot & \uline{67.47} & \uline{56.37} & \uline{47.61} \\
\bottomrule
\end{tabular}
}
\label{tab:RQ3} 
\end{table}

\begin{tcolorbox}[width=1.0\linewidth, title={}]
\textbf{Summary for RQ3:} Increasing the number of demonstration examples can improve the performance of our proposed approach, {\tool}, to some extent. However, when the number of demonstrations exceeds 4, performance decreases due to content truncation issues. In our study, using four demonstration examples achieves the best performance.
\end{tcolorbox}

\subsection{RQ4: How do different demonstration ordering strategies affect the performance of {\tool}?}

\textbf{Approach:} Inspired by the research of Gao et al.~\cite{gao2023makes}, we analyze the impact of different ordering strategies of demonstration examples on the performance of our proposed approach {\tool}. 
We hypothesize that arranging demonstration examples according to a specific ranking plays an important role in ICL. To test this hypothesis, we consider three different ordering strategies: Similarity, Reverse Similarity, and Random. Specifically, \textbf{Similarity} means arranging the demonstration examples in ascending order of their similarity to the test sample, with the examples most similar to the test sample placed closest to the target sample in the prompt template. \textbf{Reverse Similarity} is the opposite, arranged in descending order of similarity. \textbf{Random} means randomly arranging the demonstration examples.

\textbf{Results:} The results in Table~\ref{tab:RQ4} 
show that arranging the demonstration examples in ascending order of similarity to the test sample (i.e., Similarity strategy) yields the best performance. Specifically, in terms of the MCC performance measure, {\tool} outperforms the Random strategy by 5.76 pp and the Reverse Similarity strategy by 4.37 pp. This validates our hypothesis. Additionally, we observe that the performance of our approach under the Reverse Similarity strategy is better than under the Random strategy, aligning with Gao et al.'s findings for the source code summarization and program synthesis tasks~\cite{gao2023makes}. This suggests that ordering demonstration examples according to certain rules can improve ICL performance, indicating the potential for exploring more complex demonstration example ordering strategies in the future.

\begin{table}[htb]

\centering
\caption{Experimental results of different demonstration ordering strategies.}
\resizebox{8.6cm}{!}
{
\begin{tabular}{lccc}
\toprule
\textbf{Strategy} & \makecell[c]{\textbf{Accuracy (\%)}} & \makecell[c]{\textbf{F1-score (\%)}} & \makecell[c]{\textbf{MCC (\%)}} \\
\midrule
Random & 73.51 & 64.69 & 57.25 \\
Reverse Similarity & \uline{74.34} & \uline{66.38} & \uline{58.64} \\
Similarity & \textbf{77.07} & \textbf{67.89} & \textbf{63.01} \\
\bottomrule
\end{tabular}
}
\label{tab:RQ4} 
\end{table}

\begin{tcolorbox}[width=1.0\linewidth, title={}]
\textbf{Summary for RQ4:} The ordering strategy of demonstration examples affects the performance of {\tool}. Arranging them in ascending order of similarity achieves the best performance in our study.
\end{tcolorbox}

\subsection{RQ5: How does the fusion ratio of syntactic similarity and lexical similarity for source code similarity affect the performance of {\tool}?}

\textbf{Approach:} 
 In our previous work~\cite{yang2022ccgir,zhao2024automatic}, we analyzed the impact of fusing syntactic and lexical similarity with different ratio settings for source code similarity in the task of smart contract code comment generation. However, for software vulnerability assessment, it is essential to design an ablation study to determine the optimal fusion ratio. To this end, we investigated 11 different ratio settings, while maintaining the other experimental settings as described in Section~\ref{sec:Implementation Details}.

\textbf{Results:} The experimental results are shown in Table~\ref{tab:RQ5}. The first row of the table represents considering only syntactic similarity, and the last row represents considering only lexical similarity. The remaining rows show data for simultaneously considering both syntactic and lexical similarity in different proportions. It can be seen that, compared to considering a single kind of similarity, considering both kinds of similarities improves the performance of the approach {\tool}. Especially, when the syntactic similarity ratio is set to 40\% and the lexical similarity weight is set to 60\%, {\tool} achieves the best performance. Specifically, under this fusion ratio setting, {\tool} improves the performance by at least 1.66 pp in terms of the MCC measure compared to other ratio settings. This indicates that considering both syntactic and lexical similarities provides a more comprehensive measure of source code similarity.

\begin{table}[htb]

\centering
\caption{Experimental results under different fusion ratios of syntactic similarity and lexical similarity of source code. Notice SynSim represents the syntactic similarity of source code, and LexiSim represents the lexical similarity of source code.}
\resizebox{8.6cm}{!}
{
\begin{tabular}{ccccc}
\toprule
\textbf{SynSim} & \textbf{LexSim} & \makecell[c]{\textbf{Accuracy (\%)}} & \makecell[c]{\textbf{F1-score (\%)}} & \makecell[c]{\textbf{MCC (\%)}} \\
\midrule
100\% & 0\% & 71.94 & 63.81 & 53.90 \\
90\% & 10\% & 73.09 & 65.14 & 56.17 \\
80\% & 20\% & 74.34 & 66.07 & 58.35 \\
70\% & 30\% & 75.83 & 66.89 & 61.03 \\
60\% & 40\% & 74.50 & 65.81 & 58.66 \\
50\% & 50\% & 75.25 & 67.09 & 60.13 \\
40\% & 60\% & \textbf{77.07} & \textbf{67.90} & \textbf{63.01} \\
30\% & 70\% & 75.66 & 66.70 & 60.68 \\
20\% & 80\% & \uline{76.08} & \uline{67.28} & \uline{61.35} \\
10\% & 90\% & 72.10 & 63.76 & 54.74 \\
0\% & 100\% & 68.54 & 59.27 & 49.78 \\
\bottomrule
\end{tabular}
}
\label{tab:RQ5} 
\end{table}

\begin{tcolorbox}[width=1.0\linewidth, title={}]
\textbf{Summary for RQ5:} Considering both syntactic and lexical similarity can provide a comprehensive measure for source code similarity. {\tool} achieves the best performance when the ratio of syntactic similarity and lexical similarity is set as 40\% and 60\% in our study.
\end{tcolorbox}

\section{Discussion}
\label{sec:Discussion}

\subsection{\gao{Comparison with  Other Popular LLMs}}

In Section~\ref{sec:Implementation Details}, we show the reasons for using DeepSeek-V2 as our LLM. To show the effectiveness of using this LLM for {\tool}, 
\gao{we also consider four popular LLMs (i.e., GPT-3.5 Turbo, GPT-4o, GPT-4o mini, and Qwen2.5-Coder-32B-Instruct).} These LLMs have been recently used for improving software engineering tasks (such as code summarization~\cite{gao2023makes,zhao2024automatic,ahmed2024automatic}, 
bug fixing~\cite {ishizue2024improved,do2023using}, and program synthesis~\cite{zhuo2023large,li2024deveval}). In this experiment, we use the same experimental setting shown in Section~\ref{sec:Implementation Details} to guarantee a fair comparison.

\begin{table}[htb]

\centering
\caption{Performance comparison between DeepSeek-V2 and other popular LLMs.}
\resizebox{8.6cm}{!}
{
\begin{tabular}{lccc}
\toprule
\textbf{LLM} & \makecell[c]{\textbf{Accuracy (\%)}} & \makecell[c]{\textbf{F1-score (\%)}} & \makecell[c]{\textbf{MCC (\%)}} \\
\midrule
GPT-3.5 Turbo & 61.01 & 54.38 & 40.10 \\
\gao{GPT-4o mini} & 67.23 & 61.42 & 51.03 \\
GPT-4o & \uline{76.74} & \textbf{69.90} & \textbf{63.40} \\
\gao{Qwen2.5-Coder-32B-Instruct} & 64.21 & 55.93 & 48.75 \\
DeepSeek-V2 & \textbf{77.07} & \uline{67.89} & \uline{63.01} \\
\bottomrule
\end{tabular}
}
\label{tab:ComparisonwithChatGPT} 
\end{table}

\gao{As shown in Table~\ref{tab:ComparisonwithChatGPT}, GPT-3.5 Turbo achieves the worst performance among these LLMs, likely due to its context window limitation of 16k tokens, whereas the other LLMs support at least 32k tokens. Additionally, officially released data~\cite{deepseekv2} indicates that DeepSeek-V2 outperforms GPT-3.5 Turbo in areas such as mathematics, reasoning, and programming. GPT-4o mini and Qwen2.5-Coder-32B-Instruct perform better than GPT-3.5 Turbo but are still slightly inferior to DeepSeek-V2. GPT-4o mini benefits from its relatively larger context window compared to GPT-3.5 Turbo, which allows it to handle more context information. This enables it to capture more details from the source code and vulnerability descriptions. Qwen2.5-Coder-32B-Instruct, on the other hand, has shown strong performance in coding tasks. Its design and pre-training make it better at understanding code-related semantics.  However, compared to DeepSeek-V2, GPT-4o mini may lack some specific optimizations for software vulnerability assessment, and Qwen2.5-Coder-32B-Instruct may not be proficient in integrating code and vulnerability description information due to its relatively small parameter size, resulting in a slight decrease in performance. From the table, we find that GPT-4o performs slightly better than DeepSeek-V2. Given that the specific parameters and design details of GPT-4o have not yet been made public, we hypothesize that its advantage may stem from its support for an extended context of 128k tokens and its superior code analysis and language comprehension capabilities. Although GPT-4o performs slightly better than DeepSeek-V2, its cost is significantly higher. For instance, GPT-4o costs five dollars per million tokens, whereas DeepSeek-V2 costs only fourteen cents for the same number of tokens, making GPT-4o approximately 35 times more expensive. Therefore, we primarily consider DeepSeek-V2 as our LLM for {\tool}.}

\gao{In summary, these results not only confirm the competitiveness of using DeepSeek-V2 in {\tool} but also show that {\tool} can achieve relatively good performance for different LLMs when considering in-context learning and information fusion.}

\subsection{\gao{The Influence of Demonstration Similarity on Assessment Performance}}

\gao{We retrieve the specific similarity values of the selected demonstrations in our empirical study. To better illustrate the relationship between similarity values and assessment performance, we divide the test set based on the average similarity values of the demonstration examples. Note that we set the optimal number of demonstration examples to 4, as determined through our experiments.}

\begin{table}[htb]
\centering
\caption{\gao{Influence of demonstration similarity on assessment results of {\tool}. Notice: ``Avg. Sim. Range" stands for average similarity values range, and ``No. Samp." represents the number of samples.}}
\resizebox{8.6cm}{!}
{
\begin{tabular}{ccccc}
\toprule
\textbf{Avg. Sim. Range} & \textbf{No. Samp.} & \makecell[c]{\textbf{Accuracy (\%)}} & \makecell[c]{\textbf{F1-score (\%)}} & \makecell[c]{\textbf{MCC (\%)}} \\
\midrule
$$[0, 0.2)$$ & 687 & 73.37 & 65.23 & 61.58 \\
$$[0.2, 0.5)$$ & 401 & \uline{78.24} & \uline{68.01} & \uline{64.53} \\
$$[0.5, 1]$$ & 120 & \textbf{81.05} & \textbf{70.21} & \textbf{68.13} \\

\bottomrule
\end{tabular}
}
\label{tab:sec6.2} 
\end{table}

\gao{As shown in Table~\ref{tab:sec6.2}, the similarity values of the demonstration examples have an impact on the assessment results. Generally, a higher average similarity value between the demonstration examples and the target vulnerability leads to more accurate assessment and better-performing evaluation metrics. For the target vulnerabilities with an average similarity value between 0.5 and 1, the model achieves an Accuracy of 81.05\%, F1-score of 70.21\%, and an MCC of 68.13\%. In contrast, for the target vulnerabilities with an average similarity value between 0 and 0.2, the Accuracy, F1-score, and MCC decrease to 73.37\%, 65.23\%, and 61.58\%, respectively. This trend indicates that more similar demonstration examples provide the LLM with more relevant and useful information during the in-context learning process. As a result, the model better understands the characteristics of the target vulnerability and makes more accurate assessments of its severity level.}

\gao{Regarding the threshold for similarity values, although it is challenging to determine a unified threshold for different tasks, based on our experimental results, when the average similarity values of the demonstration examples are larger than 0.5, the model can achieve relatively good performance in software vulnerability assessment. However, it should be noted that the optimal threshold may vary depending on factors such as the scale and diversity of the gathered historical vulnerabilities. Specifically, larger and more diverse vulnerabilities require a higher threshold to ensure the quality of the selected demonstration examples, while smaller and less diverse vulnerabilities need a lower threshold to include sufficient relevant examples. Therefore, we follow previous studies~\cite{gao2023makes,zhao2024automatic} by selecting the top-$k$ most similar demonstrations for in-context learning.}

\subsection{The Influence of Different Dataset Splits}

\begin{figure*}[t]
    \centering
    \begin{subfigure}[b]{0.32\textwidth}
        \centering
        \includegraphics[width=\textwidth]{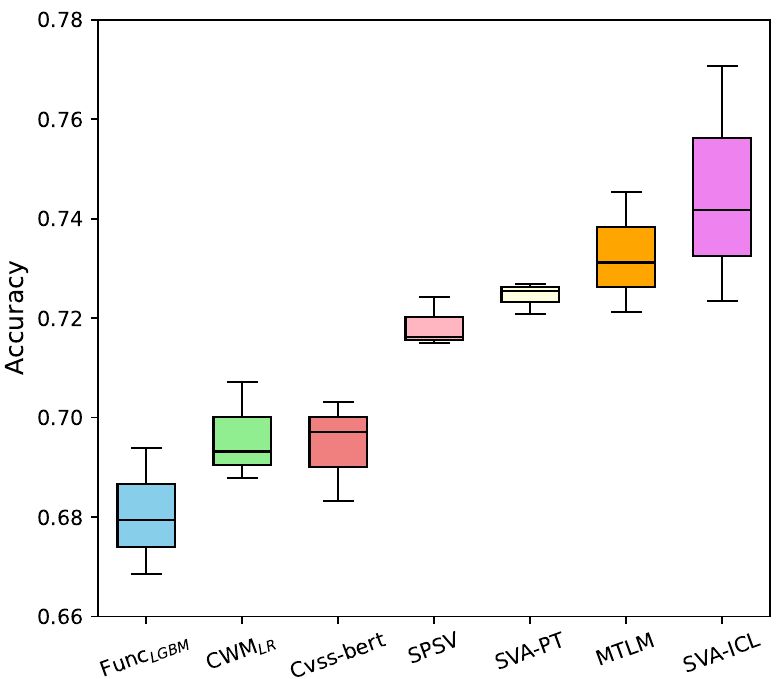}
        \caption{Accuracy}
        \label{fig:acc}
    \end{subfigure}
    \hfill
    \begin{subfigure}[b]{0.32\textwidth}
        \centering
        \includegraphics[width=\textwidth]{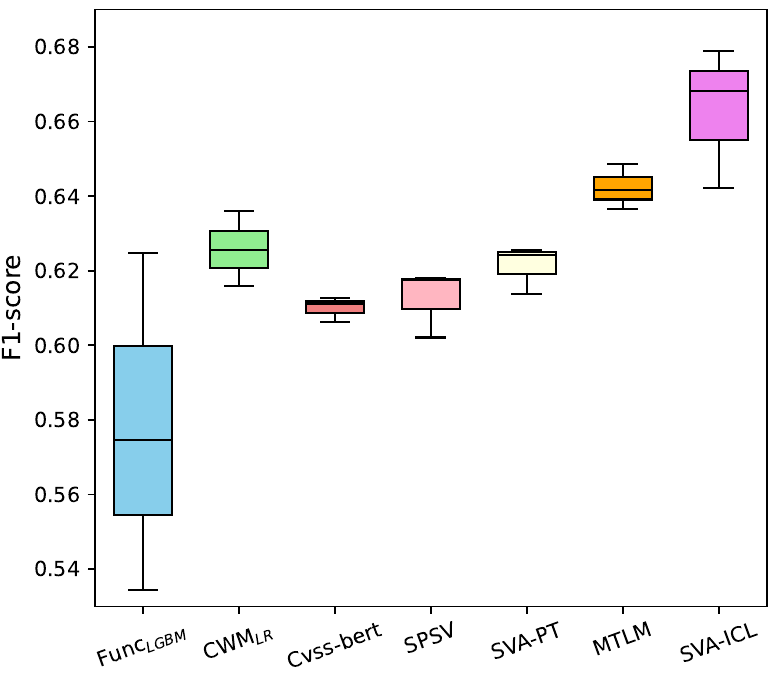}
        \caption{F1-score}
        \label{fig:f1}
    \end{subfigure}
    \hfill
    \begin{subfigure}[b]{0.32\textwidth}
        \centering
        \includegraphics[width=\textwidth]{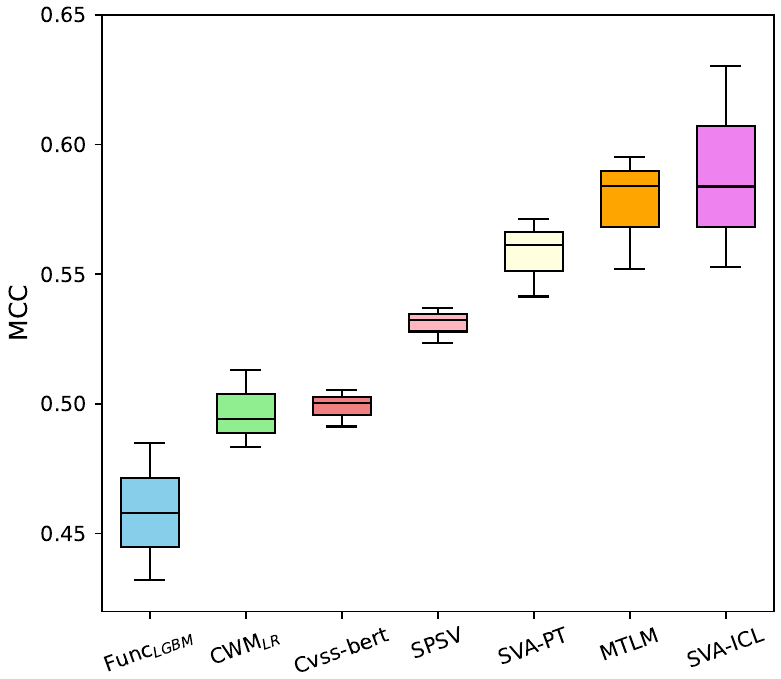}
        \caption{MCC}
        \label{fig:mcc}
    \end{subfigure}
    \caption{Comparison results between {\tool} and baselines in terms of different performance measures for different dataset splits.}
    \label{fig:random_data}
\end{figure*}

To address the inherent randomness in dataset splitting, we evaluate the performance of {\tool} and baseline approaches with different data splits. We independently split the data using various random seeds and executed these approaches under the same experimental settings described in Section~\ref{sec:Implementation Details}. \gao{The results of these three independent runs are presented in Figure~\ref{fig:random_data} using boxplots. These boxplots illustrate the performance of seven different approaches ($\text{Func}_{LGBM}$, $\text{CWM}_{LR}$, Cvss-bert, SPSV, SVA-PT, MLTM and {\tool}) in terms of Accuracy, F1-score, and MCC. Each boxplot shows the highest score (upper edge of the box), lowest score (lower edge of the box), and median score (line inside the box) achieved by these approaches over the three random splits.}

From Figure~\ref{fig:random_data}, we observe that our proposed approach {\tool} consistently achieves the best performance across all three measures (Accuracy, F1-score, MCC). \gao{Specifically, {\tool} improves Accuracy, F1-score, and MCC by an average of 3.75 pp, 4.81 pp, and 6.88 pp, respectively.} This demonstrates that the competitiveness of our proposed approach {\tool} remains robust against different data splits when compared with baseline approaches.

\gao{\subsection{Comparison with Random Demonstration Selection}}

\gao{To evaluate how the choice of demonstration examples affects model performance, we compare our relevance-based selection strategy with the random selection strategy. For a fair comparison, we set the number of selected demonstration examples to 4 for these two different strategies.  As shown in Table~\ref{tab:randomVSrelevant}, the relevance-based selection strategy achieves an accuracy of 77.07\%, significantly higher than the 61.75\% accuracy achieved by the random selection strategy. For the F1-score, the relevance-based strategy reaches 67.89\%, while the random selection strategy only achieves 46.68\%. The MCC further supports this finding, with the relevance-based strategy yielding an MCC of 63.01\%, compared to 40.92\% for the random selection strategy. In summary, the relevance-based selection strategy outperforms the random selection strategy across all three performance measures.}

\begin{table}[htb]
\centering
\caption{Experimental results of different demonstration selection strategies.}
\resizebox{8.6cm}{!}
{
\begin{tabular}{lccc}
\toprule
\textbf{Strategy} & \makecell[c]{\textbf{Accuracy (\%)}} & \makecell[c]{\textbf{F1-score (\%)}} & \makecell[c]{\textbf{MCC (\%)}} \\
\midrule
Random-based & 61.75 & 46.68 & 40.92 \\
Relevance-based & \textbf{77.07} & \textbf{67.89} & \textbf{63.01} \\
\bottomrule
\end{tabular}
}
\label{tab:randomVSrelevant}
\end{table}

\gao{\subsection{The Influence of Potential Data Leakage}}

\gao{To mitigate the potential data leakage issue, we extract test samples collected after July 2023 (i.e., the model's cutoff day for deepseek v2) and evaluate the model’s performance on this subset. We examine the performance comparison between {\tool} and the top-performing baseline, MTLM. As shown in Table~\ref{tab:dataleakage}, {\tool} outperforms MTLM across all performance measures, achieving 74.89\% accuracy (vs. 73.31\%), 65.43\% F1-score (vs. 60.79\%), and 60.68\% MCC (vs. 53.85\%).
These results show that our approach still achieves a better performance, demonstrating its effectiveness even when evaluated on data that was not seen during the pre-training phase of the LLM.}

\begin{table}[htb]
\centering
\caption{Comparison results between {\tool} and the top-performing baseline in terms of three performance measures.}
\resizebox{8.6cm}{!}
{
\begin{tabular}{lccc}
\toprule
\textbf{Approach} & \makecell[c]{\textbf{Accuracy (\%)}} & \makecell[c]{\textbf{F1-score (\%)}} & \makecell[c]{\textbf{MCC (\%)}} \\
\midrule
MTLM & 73.31 & 60.79 & 53.85 \\
{\tool} & \textbf{74.89} & \textbf{65.43} & \textbf{60.68} \\
\bottomrule
\end{tabular}
}
\label{tab:dataleakage}
\end{table}

\subsection{Insights from Our Empirical Results}

\textbf{Fusing source code and vulnerability description}.
One of the primary reasons for the superior performance of our proposed approach {\tool} is its ability to effectively integrate both source code and vulnerability descriptions through information fusion. Different from the baseline methods, which typically rely on either source code or textual descriptions alone, {\tool} combines both modalities. This dual-modality approach enhances the model's contextual understanding, allowing it to capture more comprehensive patterns that are crucial for accurate software vulnerability assessment. 
The significant drop in performance when only vulnerability descriptions were considered highlights the limitation of relying solely on textual information. Vulnerability descriptions, while informative, often lack the technical depth needed to fully understand the nature and severity of a vulnerability. Source code, on the other hand, contains critical details that directly influence the severity level, such as the specific operations and logic within the code. Therefore, the reliance on source code is necessary for accurate and precise assessments. While source code provides the foundational technical details, the integration of vulnerability descriptions appears to enhance the model’s ability to interpret and contextualize these details. Descriptions often summarize the implications or nature of the vulnerability in a way that complements the raw data provided by the code. This combination enables the model to not only identify the severity but also to understand the broader context, leading to more accurate predictions.

\textbf{Using ICL to further improve the performance of LLM-based approach}.
The results for RQ3 clearly show that the inclusion of demonstration examples substantially improves the performance of {\tool}, with the MCC increasing from 7.14\% in the zero-shot setting to 63.01\% in the four-shot setting. Interestingly, the best performance was achieved with four demonstration examples, beyond which the performance began to decline. This finding aligns with previous research~\cite{gao2023makes}, indicating that there is an optimal number of examples that maximizes the model’s ability to generalize. Too few examples (as in the zero-shot or one-shot scenarios) do not provide enough context for the model to effectively leverage its pre-trained knowledge, while too many examples can lead to information overload and potential truncation of critical content. These insights indicate that in designing prompts for tasks involving complex and lengthy inputs like software vulnerability assessment, it is crucial to carefully consider the number of examples included. The optimal number of examples will depend on the task and the specific context window limitations of the model being used. For future work, this suggests exploring dynamic or adaptive strategies that adjust the number of examples based on the specific input length and complexity.
Additionally, the results for RQ4 highlight the importance of carefully considering the order in which demonstration examples are presented in ICL tasks. For tasks like software vulnerability assessment, where context and relevance are crucial, ordering examples by similarity values can significantly enhance the model’s performance. This suggests that in future applications of ICL, particularly for complex tasks, a deliberate approach to example ordering can be a key factor in optimizing outcomes.

\textbf{Requiring comprehensive code analysis}.
The syntactic structure of source code provides a framework for code organization and flow control, while the lexical elements contain specific operations and data entities. Together, they offer a more comprehensive understanding of code behavior. Therefore, when these two similarities are effectively combined, the relationship between the target vulnerability code and the candidate vulnerability code can be captured more accurately. This also highlights the need for further exploration and optimization of code representation in future code-related software engineering studies by performing comprehensive code analysis.

\subsection{Threats to Validity}
In this subsection, we discuss the potential threats to the validity of our study.

\textbf{Internal Threats}. 
Large language models (LLMs) are trained on extensive open-source datasets, raising concerns about data leakage. This means that LLMs might have learned answers related to the test set, leading them to memorize results rather than predict them. However, our experimental results show that zero-shot learning performance is very poor, suggesting that the likelihood of data leakage is very low.

\textbf{External Threats}. 
The primary external threat is the version of the LLM API used, as different API versions may exhibit performance variations. We cannot guarantee that all versions of the API will achieve the results presented in our study. To ensure the reproducibility of our results, we recommend using the DeepSeek-V2 version of the API when invoking the LLM. Additionally, for vulnerability types that have a limited presence in the historical data, our proposed approach may not always achieve promising assessment results, but this is a common issue with data-driven approaches. To alleviate this threat, our approach is designed to be continuously updated as new data becomes available. This means that as new vulnerability types emerge, they can be incorporated into the model's training data, thereby improving its performance over time.

\textbf{Construct Threats:} 
Construct threats relate to the performance measures used. To mitigate this threat, we selected three commonly used measures in the field of software vulnerability assessment~\cite{le2022use,le2019automated,le2021deepcva}. Additionally, the construction of the prompt template is another construct threat. To address this, we followed previous research~\cite{gao2023makes} and employed carefully designed prompt templates. Our experimental results demonstrate the effectiveness of these prompt templates.

\textbf{Conclusion Threats:} 
Our constructed dataset focuses exclusively on C/C++ source code. However, since LLMs are trained on extensive datasets, our proposed approach {\tool} can also be applied to software vulnerability assessment for other programming languages. Additionally, we did not select other approaches from related work as baselines because their code is not publicly available, and implementing these approaches would be challenging.

\section{Related Work}
\label{sec:Related Work}

Software vulnerability assessment (SVA) predicts the severity of vulnerabilities based on vulnerability information (e.g., source code) to identify more critical software vulnerabilities. Developers can prioritize addressing these severe vulnerabilities based on the assessment results and propose corresponding remediation suggestions, thereby improving the overall quality of the software. In recent years, SVA research can be broadly categorized into two types based on the modalities considered: vulnerability description-based SVA and source code-based SVA.

\textbf{Vulnerability description-based SVA} uses textual descriptions of vulnerabilities to predict their severity. Han et al.~\cite{han2017learning} utilized word embeddings and a shallow convolutional neural network to capture discriminative words and sentence features in vulnerability descriptions for severity prediction. Spanos et al.~\cite{spanos2018multi} developed a model combining text analysis and multi-objective classification techniques to estimate vulnerability features and calculate severity scores. Liu et al.~\cite{liu2019vulnerability} proposed deep learning methods for vulnerability text classification, though these methods require significant time for model training. Le et al.~\cite{le2019automated} combined character and word features in their systematic approach, which necessitated tuning numerous hyperparameters. Babalau et al.~\cite{babalau2021severity} used a multi-task learning architecture with a pre-trained BERT model to compute vector-space representations of words in vulnerability descriptions, but this approach requires additional experiments to determine the values of hyperparameters and loss weights. \gao{Shahid et al.~\cite{shahid2021cvss} leveraged recent advances in the field of NLP to predict vulnerability severity scores from textual descriptions via multiple trained BERT classifiers and used a gradient-based method for explainability. }

\textbf{Source code-based SVA} involves analyzing the source code to predict potential security vulnerabilities. For instance, Ganesh et al.~\cite{ganesh2021predicting} evaluated the effectiveness of machine learning algorithms in predicting security vulnerabilities by analyzing system source code, but found that their machine learning models exhibited low performance in predictions. Le et al.~\cite{le2022use} investigated machine learning models for automatically performing function-level SVA and explored the value of using vulnerable statements as inputs for developing assessment models. Hao et al.~\cite{hao2023novel} proposed a source code-level vulnerability severity assessment approach that does not require vulnerability reports or manual analysis, allowing for quick severity assessment by combining function call graphs and vulnerability attribute graphs.

\gao{\textbf{Bimodal data-based SVA} uses both vulnerability description and source code to predict severity. Xue et al.~\cite{xue2025towards} proposed a method for SVA, applying prompt tuning and continual learning. It combined confidence-based replay and regularization, used source code and vulnerability descriptions to create hybrid prompts for tuning with CodeT5. Du et al.~\cite{du2024vulnerability} proposed an SVA method based on bimodal data and multi-task learning. They preprocessed bimodal data (description and source code), used GraphCodeBert for feature extraction, Bi-GRU with attention for further extraction, and a multi-task learning approach with hard parameter sharing.}

Different from the previous research, we are the first to introduce large language models into the task of software vulnerability assessment and utilize in-context learning to enhance the performance of vulnerability evaluation. To select high-quality demonstration examples, we apply an information fusion approach that comprehensively considers both source code and vulnerability descriptions. Additionally, we analyze the impact of the number of examples, ordering strategies, and code similarity measurement methods on the performance of {\tool}.

\section{Conclusion and Future Work}
\label{sec:Conclusion}

In this study, we propose a novel approach {\tool}, which leverages the in-context learning capabilities of large language models for software vulnerability assessment. By performing information fusion for source code and vulnerability description, {\tool} can select high-quality demonstration examples from historical corpora. These examples are used to construct customized prompts during the ICL stage, which can aid the LLM DeepSeek-V2 in leveraging relevant knowledge for software vulnerability assessment. Experimental results on our constructed dataset demonstrate that {\tool} outperforms SVA baselines in terms of various performance measures. Finally, we conduct a set of ablation studies to show the effectiveness of the customization of {\tool}, such as the demonstration number, and the ordering strategy.

Despite the promising performance of our {\tool} approach in software vulnerability assessment, there are still several directions worth further exploration. First, we want to design more efficient demonstration example selection strategies to further improve the performance of {\tool}. Second, we want to design more high-quality prompts to guide LLM to improve the assessment performance. Lastly, we want to apply {\tool} to other programming languages to show the generalization of {\tool}.

\section*{CRediT authorship contribution statement}

\textbf{Chaoyang Gao:} Data curation, Software, Validation, Conceptualization, Methodology, Writing -review \& editing.
\textbf{Xiang Chen:} Conceptualization, Methodology, Writing -review \& editing, Supervision.

\textbf{Guangbei Zhang}: Conceptualization, Data curation, Software.

\section*{Declaration of competing interest}
The authors declare that they have no known competing financial interests or personal relationships that could have appeared to
influence the work reported in this paper.

\section*{Data availability}

We share our data on GitHub (\url{https://github.com/judeomg/SVA-ICL}).

\section*{Acknowledgments}
Chaoyang Gao and Xiang Chen have contributed equally to this work and are co-first authors. Xiang Chen is the corresponding author. 
This research was partially supported by the National Natural Science Foundation of China (Grant No. 61202006), the Open Project of State Key Laboratory for Novel Software Technology at Nanjing University under (Grant No. KFKT2024B21), and the Postgraduate Research \& Practice Innovation Program of Jiangsu Province (Grant No. SJCX24\_2022).

\bibliography{mylib}
\bibliographystyle{elsarticle}

\vspace{1cm}

\noindent\textbf{Chaoyang Gao} 
is currently pursuing the Master degree at the School of Artificial Intelligence and Computer Science, Nantong University. His research interests include software
repository mining.
\par
\vspace{1cm}

\noindent\textbf{Xiang Chen} 
  received the B.Sc. degree in information management and systems from Xi'an Jiaotong University, China in 2002. Then he received his M.Sc., and Ph.D. degrees in computer software and theory from Nanjing University in 2008 and 2011, respectively. He is an Associate Professor at the School of Artificial Intelligence and Computer Science, Nantong University. He has authored or co-authored more than 160 papers in refereed journals or conferences, such as IEEE Transactions on Software Engineering, ACM Transactions on Software Engineering and Methodology, Empirical Software Engineering, Information and Software Technology, Journal of Systems and Software, Software Testing, Verification and Reliability, Journal of Software: Evolution and Process, International Conference on Software Engineering (ICSE), International Conference on the Foundations of Software Engineering (FSE), International Symposium on Software Testing and Analysis (ISSTA), International Conference Automated Software Engineering (ASE), International Conference on Software Maintenance and Evolution (ICSME), International Conference on Program Comprehension (ICPC), and International Conference on Software Analysis, Evolution and Reengineering (SANER). His research interests include software engineering, in particular software testing and maintenance, software repository mining, and empirical software engineering. He received two ACM SIGSOFT distinguished paper awards in ICSE 2021 and ICPC 2023. He is the editorial board member of Information and Software Technology. More information can be found at:

    https://xchencs.github.io/index.html.

\par
\vspace{1cm}

\noindent\textbf{Guangbei Zhang} 
is currently pursuing the Bachelor degree at the School of Artificial Intelligence and Computer Science, Nantong University. Her research interests include automatic vulnerability assessment.

\end{document}